\documentclass[a4paper,11pt]{article}
\usepackage{jinstpub}

\usepackage{lineno}

\usepackage[utf8]{inputenc}
\usepackage{array}
\usepackage{makecell}
\usepackage{outlines}
\usepackage{ulem}
\usepackage{siunitx}
\usepackage{tikz}

\title{\boldmath The Skin Detector System of the LZ Dark Matter Experiment}

\author[1,2]{D.S.~Akerib}
\author[3]{A.K.~Al Musalhi}
\author[3]{F.~Alder}
\author[4]{B.J.~Almquist}
\author[5]{C.S.~Amarasinghe}
\author[1,2]{A.~Ames}
\author[6]{T.J.~Anderson}
\author[7]{N.~Angelides}
\author[8,9]{H.M.~Ara\'{u}jo}
\author[10]{J.E.~Armstrong}
\author[1,2]{M.~Arthurs}
\author[11]{A.~Baker}
\author[9]{S.~Balashov}
\author[4]{J.~Bang}
\author[5]{J.W.~Bargemann}
\author[7]{E.E.~Barillier}
\author[6]{K.~Beattie}
\author[12]{A.~Beauchene}
\author[10]{A.~Bhatti}
\author[1,2]{T.P.~Biesiadzinski}
\author[7]{H.J.~Birch}
\author[13]{E.~Bishop}
\author[14]{G.M.~Blockinger}
\author[9]{C.A.J.~Brew}
\author[15]{P.~Br\'{a}s}
\author[16]{S.~Burdin}
\author[17]{M.C.~Carmona-Benitez}
\author[16]{M.~Carter}
\author[18]{A.~Chawla}
\author[17]{Y.T.~Chin}
\author[19]{S.~Contreras}
\author[20]{M.V.~Converse}
\author[1,2]{R.~Coronel}
\author[3]{A.~Cottle}
\author[21]{G.~Cox}
\author[21]{D.~Curran}
\author[22,23]{C.E.~Dahl}
\author[3]{I.~Darlington}
\author[3]{S.~Dave}
\author[3]{A.~David}
\author[21]{J.~Delgaudio}
\author[12]{S.~Dey}
\author[17]{L.~de~Viveiros}
\author[8]{L.~Di Felice}
\author[4]{C.~Ding}
\author[11]{J.E.Y.~Dobson}
\author[20]{E.~Druszkiewicz}
\author[4]{S.~Dubey}
\author[21]{C.L.~Dunbar}
\author[24]{S.R.~Eriksen}
\author[12]{N.M.~Fearon}
\author[12]{N.~Fieldhouse}
\author[6]{S.~Fiorucci}
\author[24]{H.~Flaecher}
\author[16]{E.D.~Fraser}
\author[25]{T.M.A.~Fruth}
\author[1,2]{P.W.~Gaemers}
\author[4]{R.J.~Gaitskell}
\author[21]{A.~Geffre}
\author[17,26]{J.~Genovesi}
\author[3]{C.~Ghag}
\author[11]{J.~Ghamsari}
\author[14]{A.~Ghosh}
\author[1,2]{S.~Ghosh}
\author[27,6]{R.~Gibbons}
\author[28]{S.~Gokhale}
\author[3]{J.~Green}
\author[9]{M.G.D.~van~der~Grinten}
\author[26]{J.J.~Haiston}
\author[10]{C.R.~Hall}
\author[16]{T.~Hall}
\author[7]{R.N.~Hampp}
\author[29]{S.J.~Haselschwardt}
\author[7]{M.A.~Hernandez}
\author[30]{S.A.~Hertel}
\author[31]{G.J.~Homenides}
\author[21]{M.~Horn}
\author[19]{D.Q.~Huang}
\author[12,32]{D.~Hunt}
\author[8]{E.~Jacquet}
\author[33]{R.~James}
\author[15]{K.~Jenkins}
\author[18]{A.C.~Kaboth}
\author[19]{A.C.~Kamaha}
\author[14]{M.K.~Kannichankandy  }
\author[20]{D.~Khaitan}
\author[9]{A.~Khazov}
\author[5]{J.~Kim} \emailAdd{jeonghwa\_kim@ucsb.edu}
\author[34]{Y.D.~Kim}
\author[6]{D.~Kodroff }
\author[35]{E.V.~Korolkova}
\author[12]{H.~Kraus}
\author[32]{S.~Kravitz}
\author[24]{L.~Kreczko}
\author[35]{V.A.~Kudryavtsev}

\author[5,\dagger]{T.A.~Kumar}
\author[11]{C.~Lawes}
\author[29]{E.B.~Leon}
% ...

\affiliation[1]{SLAC National Accelerator Laboratory, Menlo Park, CA 94025-7015, USA}

\affiliation[2]{Kavli Institute for Particle Astrophysics and Cosmology, Stanford University, Stanford, CA  94305-4085 USA}

\affiliation[3]{University College London (UCL), Department of Physics and Astronomy, London WC1E 6BT, UK}

\affiliation[4]{Brown University, Department of Physics, Providence, RI 02912-9037, USA}

\affiliation[5]{University of California, Santa Barbara, Department of Physics, Santa Barbara, CA 93106-9530, USA}

\affiliation[6]{Lawrence Berkeley National Laboratory (LBNL), Berkeley, CA 94720-8099, USA}

\affiliation[7]{University of Zurich, Department of Physics, 8057 Zurich, Switzerland}

\affiliation[8]{Imperial College London, Physics Department, Blackett Laboratory, London SW7 2AZ, UK}

\affiliation[9]{STFC Rutherford Appleton Laboratory (RAL), Didcot, OX11 0QX, UK}

\affiliation[10]{University of Maryland, Department of Physics, College Park, MD 20742-4111, USA}

\affiliation[11]{King's College London, King's College London, Department of Physics, London WC2R 2LS, UK}

\affiliation[12]{University of Oxford, Department of Physics, Oxford OX1 3RH, UK}

\affiliation[13]{University of Edinburgh, SUPA, School of Physics and Astronomy, Edinburgh EH9 3FD, UK}

\affiliation[14]{University at Albany (SUNY), Department of Physics, Albany, NY 12222-0100, USA}

\affiliation[15]{{Laborat\'orio de Instrumenta\c c\~ao e F\'isica Experimental de Part\'iculas (LIP)}, University of Coimbra, P-3004 516 Coimbra, Portugal}

\affiliation[16]{University of Liverpool, Department of Physics, Liverpool L69 7ZE, UK}

\affiliation[17]{Pennsylvania State University, Department of Physics, University Park, PA 16802-6300, USA}

\affiliation[18]{Royal Holloway, University of London, Department of Physics, Egham, TW20 0EX, UK}

\affiliation[19]{University of California, Los Angeles, Department of Physics \& Astronomy, Los Angeles, CA 90095-1547}

\affiliation[20]{University of Rochester, Department of Physics and Astronomy, Rochester, NY 14627-0171, USA}

\affiliation[21]{South Dakota Science and Technology Authority (SDSTA), Sanford Underground Research Facility, Lead, SD 57754-1700, USA}

\affiliation[22]{Northwestern University, Department of Physics \& Astronomy, Evanston, IL 60208-3112, USA}

\affiliation[23]{Fermi National Accelerator Laboratory (FNAL), Batavia, IL 60510-5011, USA}

\affiliation[24]{University of Bristol, H.H. Wills Physics Laboratory, Bristol, BS8 1TL, UK}

\affiliation[25]{The University of Sydney, School of Physics, Physics Road, Camperdown, Sydney, NSW 2006, Australia}

\affiliation[26]{South Dakota School of Mines and Technology, Rapid City, SD 57701-3901, USA}

\affiliation[27]{University of California, Berkeley, Department of Physics, Berkeley, CA 94720-7300, USA}

\affiliation[28]{Brookhaven National Laboratory (BNL), Upton, NY 11973-5000, USA}

\affiliation[29]{University of Michigan, Randall Laboratory of Physics, Ann Arbor, MI 48109-1040, USA}

\affiliation[30]{University of Massachusetts, Department of Physics, Amherst, MA 01003-9337, USA}

\affiliation[31]{University of Alabama, Department of Physics \& Astronomy, Tuscaloosa, AL 34587-0324, USA}

\affiliation[32]{University of Texas at Austin, Department of Physics, Austin, TX 78712-1192, USA}

\affiliation[33]{University of Melbourne, School of Physics, Melbourne, VIC 3010, Australia}

\affiliation[34]{IBS Center for Underground Physics (CUP), Yuseong-gu, Daejeon, Korea}

\affiliation[35]{University of Sheffield, School of Mathematical and Physical Sciences, Sheffield S3 7RH, UK}

% ...
\affiliation[\dagger]{Now at Perimeter Institute for Theoretical Physics, Waterloo, ON N2L 2Y5, Canada}

\abstract{
This paper presents the design, calibration, and performance of the Skin Detector component of the LUX-ZEPLIN (LZ) dark matter experiment. The Skin Detector consists of approximately 2 tonnes of liquid xenon instrumented with 131 photomultiplier tubes, surrounding the central time-projection chamber and functions as an active veto system for background rejection. Energy calibration with monoenergetic gamma sources established a position-dependent energy scale and determined prompt and delayed veto thresholds. Radon decays in the Skin can be identified with less than 15~mm spatial resolution, revealing stable convective flow patterns with measured velocities ranging from 2-4 mm/s that persisted throughout extended data-taking periods. The achieved Skin veto threshold meets or exceeds the design goal of 100~keV throughout the Skin volume. The prompt veto efficiency for gamma rays emitted following electron capture in $^{127}$Xe is found to be ($88 \pm 2$)\%, consistent with the simulated value of ($90.4 \pm 0.6$)\%. M-shell EC decays are used to validate the LZ detector sensitivity for low-mass dark matter during the first science run.

}

\keywords{}

\arxivnumber{}

\begin{document}
\maketitle
\flushbottom

\section{Introduction}\label{sec:Intro}

\begin{figure}
	\centering
    \includegraphics[width=0.9\linewidth]{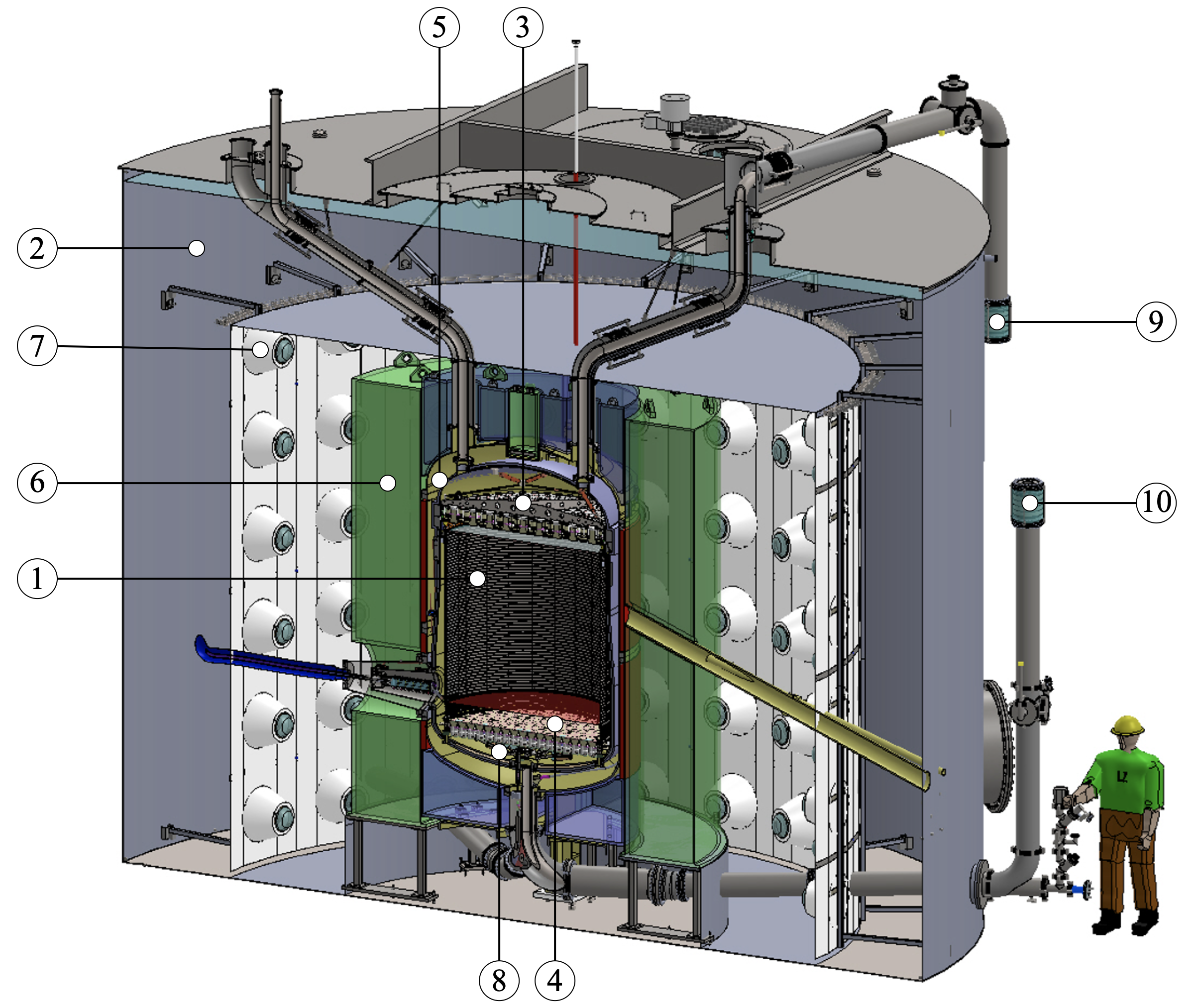}
	\caption{Overview of the LZ detector system with major subsystems labeled. The central TPC (1) is immersed in liquid xenon and enclosed within a water tank (2) that provides passive shielding. Scintillation signals are collected by two PMT arrays: the top (3) and bottom (4) TPC arrays. The TPC is housed inside a double-walled vacuum-insulated titanium cryostat (5) and optically isolated from the surrounding Skin and Outer Detector. A gadolinium-loaded liquid scintillator (6) surrounds the cryostat and is viewed by an array of inward-facing PMTs (7). The Skin Detector PMTs are installed in two regions: the Barrel around the vertical walls of the TPC and the Dome beneath it (8). Signal and high-voltage cables for the TPC and Skin are routed via top (9) and bottom (10) penetrations leading to the breakout region, where analog electronics are located.}
	\label{fig:LZModel}
\end{figure}

\begin{figure}
	\centering
    \includegraphics[width=0.9\linewidth]{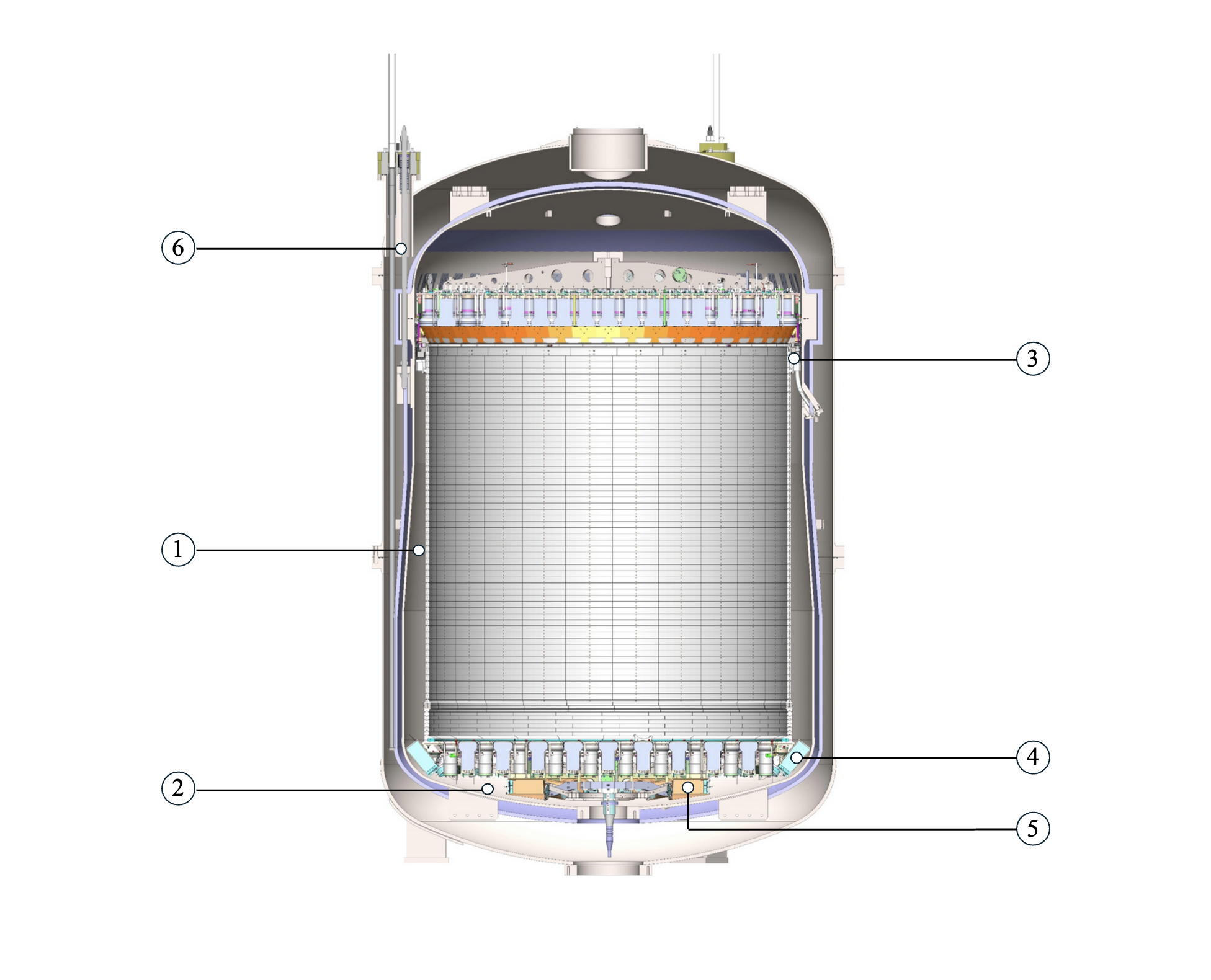}
	\caption{Cutaway CAD rendering showing the cryostat (outer and inner cryostat vessel), the TPC field cage, and the Skin Detector volume outside of the field cage. The Skin volume is divided into the Barrel (1) surrounding the vertical walls of the TPC field cage, and the Dome (2) beneath the TPC. Scintillation light in the Skin is detected by the top Skin PMTs (3) mounted above the TPC, bottom Skin PMTs (4) positioned at the base of the Barrel region, and Dome PMTs (5) viewing the underside of the TPC from below. The calibration source deployment (CSD) tube (6) runs vertically through the detector.}
	\label{fig:SkinModel}
\end{figure}

In this article, we describe the design and performance of the LUX-ZEPLIN (LZ) Skin Detector. The LZ experiment~\cite{akerib2020lux} is a  dark matter direct detection experiment designed to achieve unprecedented sensitivity to weakly interacting massive particles (WIMPs), one of the leading candidates for dark matter. Located at the 4850-foot level of the Sanford Underground Research Facility (SURF) in Lead, South Dakota, USA, LZ utilizes a dual-phase xenon time projection chamber (TPC) to detect nuclear recoils from WIMP interactions. To date, LZ has published results from three science runs, referred to as WS2022~\cite{aalbers2023first}, WS2024~\cite{aalbers2025dark}, and WS2025~\cite{akerib2025searches}, with world-leading constraints on spin-independent and spin-dependent WIMP-nucleon cross sections.
Figure~\ref{fig:LZModel} shows the layout of the LZ experiment.

Ionizing interactions in the TPC produce two distinct signals: a prompt scintillation pulse (S1) and a delayed electroluminescence pulse (S2) from ionization electrons extracted into the gas phase. Voltages applied to three wire-mesh grid electrodes (cathode, gate, and anode) create the electric field configuration necessary for electron drift, extraction and electroluminescence. The S1 and S2 signals are detected by two arrays of photomultiplier tubes (PMTs) located above and below the LXe TPC volume. The timing and spatial distribution of S1 and S2 allow for three-dimensional event reconstruction, and their relative sizes allow discrimination between nuclear and electronic recoils. However, given the extreme rarity of potential WIMP interactions, additional background suppression is essential. To this end, LZ employs both passive shielding and two active veto systems, the xenon Skin Detector and the Outer Detector (OD).

The Skin Detector, regions 1 and 2 in Fig.~\ref{fig:SkinModel}, refers to the instrumented LXe volume outside the TPC field cage but within the inner cryostat vessel (ICV). The TPC electrodes and the grounded ICV create a horizontal electric field in the Skin volume, approximately within the range 3--6 kV/cm in WS2022 and from 1.5--2 kV/cm in WS2024 and WS2025. With no electroluminescence region, the Skin Detector produces only S1-like signals. 

The Skin Detector serves a dual purpose: electrically, it provides the necessary standoff between the biased TPC components and the grounded ICV; functionally, it acts as an active veto for background events. A gamma-ray that Compton scatters in the TPC and then deposits additional energy in the Skin will produce a coincident signal that is inconsistent with the expected single-scatter topology of a WIMP interaction. Similarly, neutrons that scatter in the TPC and are subsequently captured in the Skin or OD can be identified through delayed coincidences. By rejecting such events, the Skin significantly enhances the background rejection capability of the experiment, especially for events near the edge of the detector volume where external backgrounds are more likely to penetrate. 

In addition to its veto function, the Skin Detector serves as a diagnostic tool for characterizing the background environment of the experiment. With sufficient light collection and timing resolution, the Skin enables energy and position reconstruction of interactions within its volume. This allows for the identification of radioactive contaminants, such as gamma rays from detector materials, neutron captures, and alpha decays from radon progeny. Notably, the Skin has demonstrated sensitivity to decay chains from $^{222}$Rn and $^{220}$Rn, enabling spatial mapping of the liquid flow by tracking the decay pairs. The Skin has also been used to tag electron capture decays of xenon isotopes such as $^{127}$Xe, which produce low energy Auger electrons and X-rays with associated gamma-ray emission~\cite{aalbers2025measurements}.

This paper presents a study of the LZ Skin Detector, including its hardware configuration, energy calibration, position resolution, and veto performance. Section 2 describes the detector geometry, PMT configuration, and readout system. Section 3 details the methods used to calibrate the Skin position and energy scales using external and internal gamma-ray sources. Section 4 presents the reconstruction of internal alpha emitters, position reconstruction performance, and mapping of the liquid flow. Section 5 evaluates the Skin’s veto capabilities for prompt and delayed signals, comparing measured performance with design requirements.

\section{Detector Description}\label{sec:Hardware}

\subsection{Detector Geometry}

The Skin Detector is a single-phase LXe scintillator volume that envelops the TPC~\cite{akerib2020lux}, providing an active veto for any interactions producing energy deposits in the Skin volume (Fig.~\ref{fig:SkinModel}). It is optically and electrically separated from the TPC and is divided into two regions: a cylindrical "Barrel" surrounding the vertical walls of the TPC and a "Dome" located beneath the bottom PMT array. Together, these regions contain approximately 2 tonnes of LXe and form a contiguous volume between the TPC field cage and the ICV.

The Skin's Barrel has an azimuthally symmetric geometry, with its radial thickness varying from 4 cm near the liquid surface to 8 cm at the bottom, due to the tapered shape of the ICV~\cite{mount2017lux}. The vertical extent of approximately 160 cm covers the liquid volume from just below the liquid surface to the bottom TPC PMT array. The Skin's Dome, with an outer radius of 82 cm matching the lower parts of the Barrel, occupies the space below the bottom TPC PMT array above the bottom of the inner cryostat. The Dome varies in depth from a few cm to a maximum depth of about 12 cm. Calibration Source Deployment (CSD) tubes~\cite{aalbers2024design} located between the inner and outer cryostat vessels allow source deployment near the Skin and Outer Detector at azimuthal angles of $\theta$ = 0$^\circ$, 120$^\circ$, and 240$^\circ$, reaching down to the cathode.

Both Skin regions are lined with high-reflectivity polytetrafluoroethylene (PTFE) to enhance light collection efficiency for vacuum ultraviolet (VUV) scintillation light. The PTFE used is Technetics 8764, made from Daikin M17 powder, which was selected after an R\&D campaign for its combination of high VUV reflectivity of $\geq$ 97\%~\cite{Neves:2016tcw,Kravitz:2019zqv} and low intrinsic radioactivity. PTFE tiles of 1 mm thickness line the inner surface of the ICV. The tiles are held in place by PTFE washers and screws mated to low-profile, threaded titanium buttons glued to the ICV wall using cryogenic-grade MasterBond EP29LPSP epoxy. 
The tiles cover the full inner surface except the high-voltage (HV) feedthrough region. The outer surface of the TPC field cage is also made of PTFE, serving as an additional reflective boundary.

\begin{figure}
	\centering
	\includegraphics[width=0.95\linewidth]{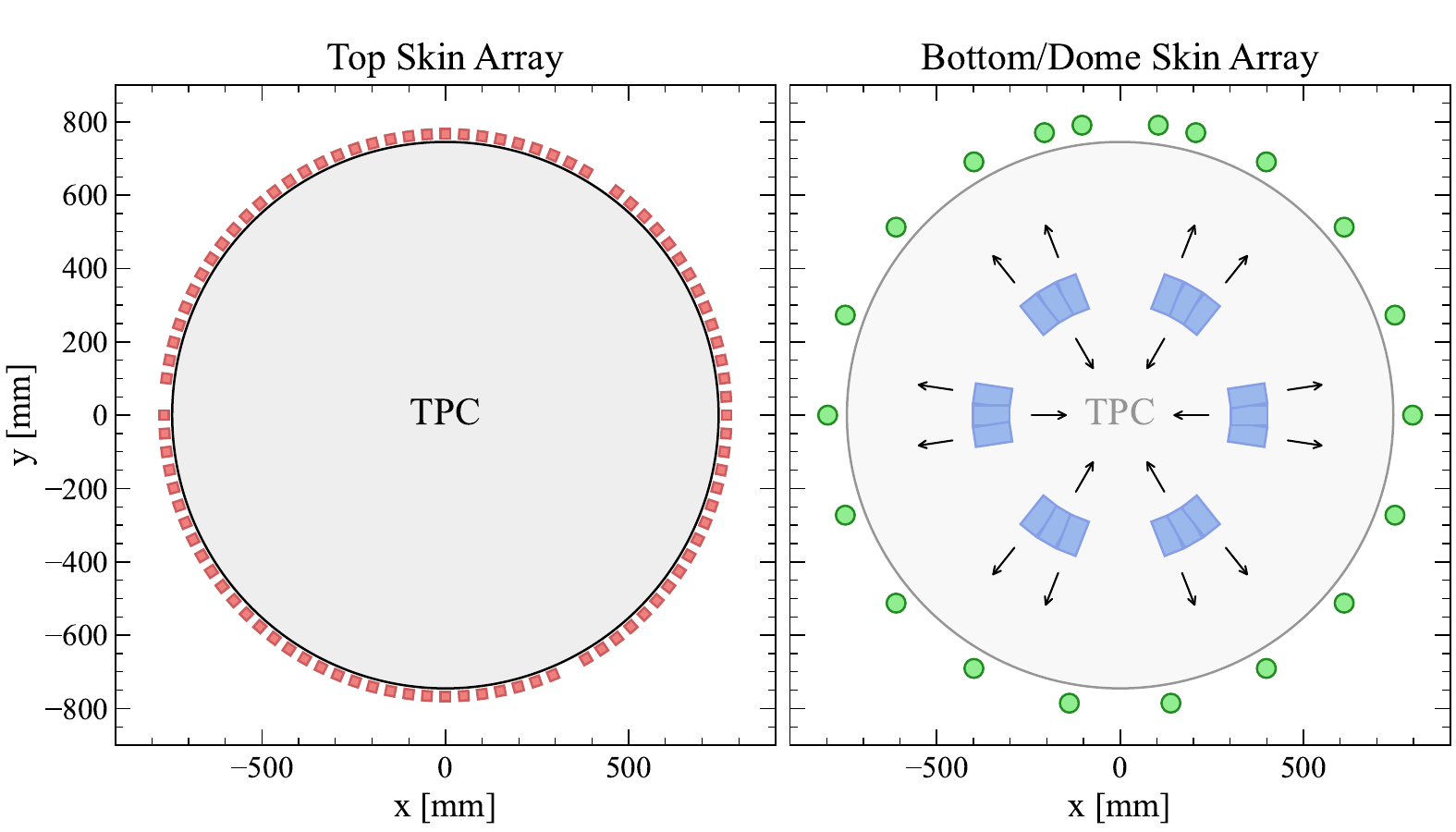}
	\caption{PMT layout in the Skin Detector. The top Skin region is equipped with Hamamatsu R8520 PMTs (red) mounted azimuthally around the Barrel, while the bottom Skin (green) and Dome (blue) regions employ R8778 PMTs arranged in vertical and concentric configurations respectively. Arrows indicate the normal incidence viewing directions of the photocathodes for Dome PMTs. Two extra bottom PMTs are directed toward the HV feedthrough region. Three small gaps in the top array are due to the weir drains.}
	\label{fig:PMT Diagram}
\end{figure}

\subsection{PMT Specifications and Configuration}

The Skin Detector is instrumented with a total of 131 PMTs as shown in Fig.~\ref{fig:PMT Diagram}, distributed across the Barrel and Dome regions to provide high-efficiency detection of scintillation light from interactions in the surrounding LXe. The PMTs were selected to balance light collection performance, mechanical constraints, and radiopurity requirements. Two models of Hamamatsu PMTs are used: 93 of the compact 1-inch-square R8520-406s, and 38 of the larger 2-inch-diameter R8778s, with the latter reused from the LUX experiment~\cite{akerib2013ultra}. 

The upper Barrel region is equipped with R8520-406 PMTs, arranged in a single layer just below the liquid xenon surface and mounted facing downward. These PMTs are attached to the outer surface of the TPC field cage. At the top of the detector, they are supported by the weir structure that sets the liquid level in the TPC. The compact form factor of the R8520-406 allows dense packing within the limited radial space available in the upper Skin, with three gaps present to allow for drainage tubes for the weir (visible in the upper right part of Fig.~\ref{fig:SkinModel}). These PMTs are optimized for xenon scintillation light collection, as they feature a quartz window and a bialkali photocathode with a quantum efficiency of ~30\% at 175 nm and a collection efficiency for electrons at the first dynode of 70\%~\cite{aprile2012measurement}.

The lower Barrel region is instrumented with 20 R8778 PMTs, positioned radially around the base of the TPC and oriented upward. Eighteen of these PMTs provide uniform coverage of the lower Barrel region, and two are specifically directed toward the HV feedthrough region, where it was not possible to attach the PTFE liner. These PMTs also feature similar characteristics with a quantum efficiency of 33\% at 175 nm and a collection efficiency of 90\%~\cite{akerib2013ultra}.

The Dome is instrumented with an additional 18 R8778 PMTs, mounted horizontally beneath the bottom PMT array. These are arranged in two concentric rings: 12 PMTs facing radially outward and 6 facing radially inward. This configuration is designed to optimize light collection efficiency in the Dome, which has a complex geometry due to the curvature of the cryostat and the presence of structural supports and cables. All R8778 PMTs as well as the bottom TPC PMT array (consisting of 241 R11410 PMTs~\cite{Akerib:2019fml}) are sleeved in PTFE to enhance reflectivity and reduce photon absorption on metal surfaces. 

The Skin includes a dedicated LED calibration system with 30 LEDs (470 nm) distributed across the detector: 12 above the upper Barrel PMTs, 12 below the lower Barrel PMTs, and 6 in the Dome region. The LEDs are housed in PTFE structures to diffuse light and provide uniform illumination. This system operates at low intensities to measure single photoelectron response for per-PMT gain calibration, and at higher intensities to characterize afterpulsing ratios as a diagnostic of PMT vacuum quality.

\subsection{Signal Chain and Data Acquisition}

The Skin PMTs are read out using the signal processing and data acquisition (DAQ) infrastructure developed for the LZ experiment~\cite{aalbers2024data}.
PMT signals are transmitted through vacuum feedthroughs located on the top and bottom cryostat flanges and then routed to analog front-end electronics housed in mini-crates attached directly to feedthroughs. Amplifier cards provide an amplification of a factor of 40 with a shaping time of 60 ns full width at one tenth maximum~\cite{aalbers2024data}. 

The amplified signals are digitized per channel, with sampling rates of 100 MHz and 14-bit resolution. The DAQ system employs a pulse-only-digitization scheme to reduce data volume, storing only waveform segments surrounding detected pulses. 

A series of trigger sources determine when data are recorded in LZ~\cite{aalbers2024data}, and Skin waveforms are recorded for all trigger types. In WS2022, the Skin itself could not trigger the detector; subsequently, a dedicated Skin trigger was implemented to enable efficient recording of radon-chain alpha decays, which are a key calibration source for the Skin Detector. The Skin trigger achieves a high efficiency of 99.5\% for these decays while maintaining a low trigger rate of approximately 0.5 Hz, ensuring compatibility with the overall LZ trigger budget of $\sim$20 Hz. 

\section{Detector Calibration}\label{sec:Calibration}

The ability to reconstruct the location and energy depositions of events in the Skin is essential for optimizing the veto performance and for characterizing background populations. While the TPC uses S2 electroluminescence signals for drift-time-based $z$-position reconstruction, the Skin does not have S2 readout and thus cannot provide this capability. However, it enables partial spatial reconstruction through the distribution of signal among its PMTs. Two principal observables are used for position inference: the azimuthal angle $\theta$, reconstructed from the weighted average of PMT hits in the cylindrical Barrel (Fig.~\ref{fig:PMT Diagram}), and the vertical coordinate $z$, inferred from the top-bottom asymmetry (TBA) of the light collected in the top and bottom/Dome PMT arrays:
\[
\mathrm{TBA} = \frac{\mathrm{S1}_{\mathrm{top}} - \mathrm{S1}_{\mathrm{bottom}}}{\mathrm{S1}_{\mathrm{top}} + \mathrm{S1}_{\mathrm{bottom}}},
\]
where $\mathrm{S1}_{\mathrm{top}}$ and $\mathrm{S1}_{\mathrm{bottom}}$ represent the summed S1 pulse areas detected by the top and bottom+Dome arrays in units of photons detected (phd).

The S1 area also serves as the primary observable for energy reconstruction in the Skin. Since alphas, betas, gamma rays, and neutrons produce differing scintillation yields relative to deposited energy~\cite{doke2002absolute}, a universal energy scale cannot be defined. Given that the primary function of the Skin is to veto gamma-induced backgrounds, the energy calibration is performed using monoenergetic gamma sources and is valid specifically for gamma interactions. Since gammas and electrons produce nearly identical scintillation yields at the electric fields present in the Skin, the reconstructed gamma energy scale is expressed in electron-equivalent energy (keV$_\text{ee}$).

External calibrations were performed using monoenergetic sources deployed through the CSD tubes following WS2022. Four gamma sources: $^{57}$Co (122 keV), $^{54}$Mn (835 keV), $^{22}$Na (511 keV and 1275 keV), and $^{228}$Th (2615 keV) were deployed at six vertical positions ranging from 0 mm to 2500 mm above the cathode, enabling a detailed mapping of the Skin Barrel. Internal backgrounds from science runs, such as $^{127}$Xe and $^{133}$Xe, are also used for position calibration.

The azimuthal symmetry of the detector allows for a straightforward $\theta$ reconstruction. In contrast, vertical position and energy reconstruction are complicated by asymmetries in the PMT arrays, electric field variations, and reflections within the thin LXe volume. The following two subsections outline the procedures to produce data-driven maps of $z$(TBA) and $E_\gamma$(S1, TBA).

\subsection{Vertical Position ($z$)}

\begin{figure}
	\centering
	\includegraphics[width=0.95\linewidth]{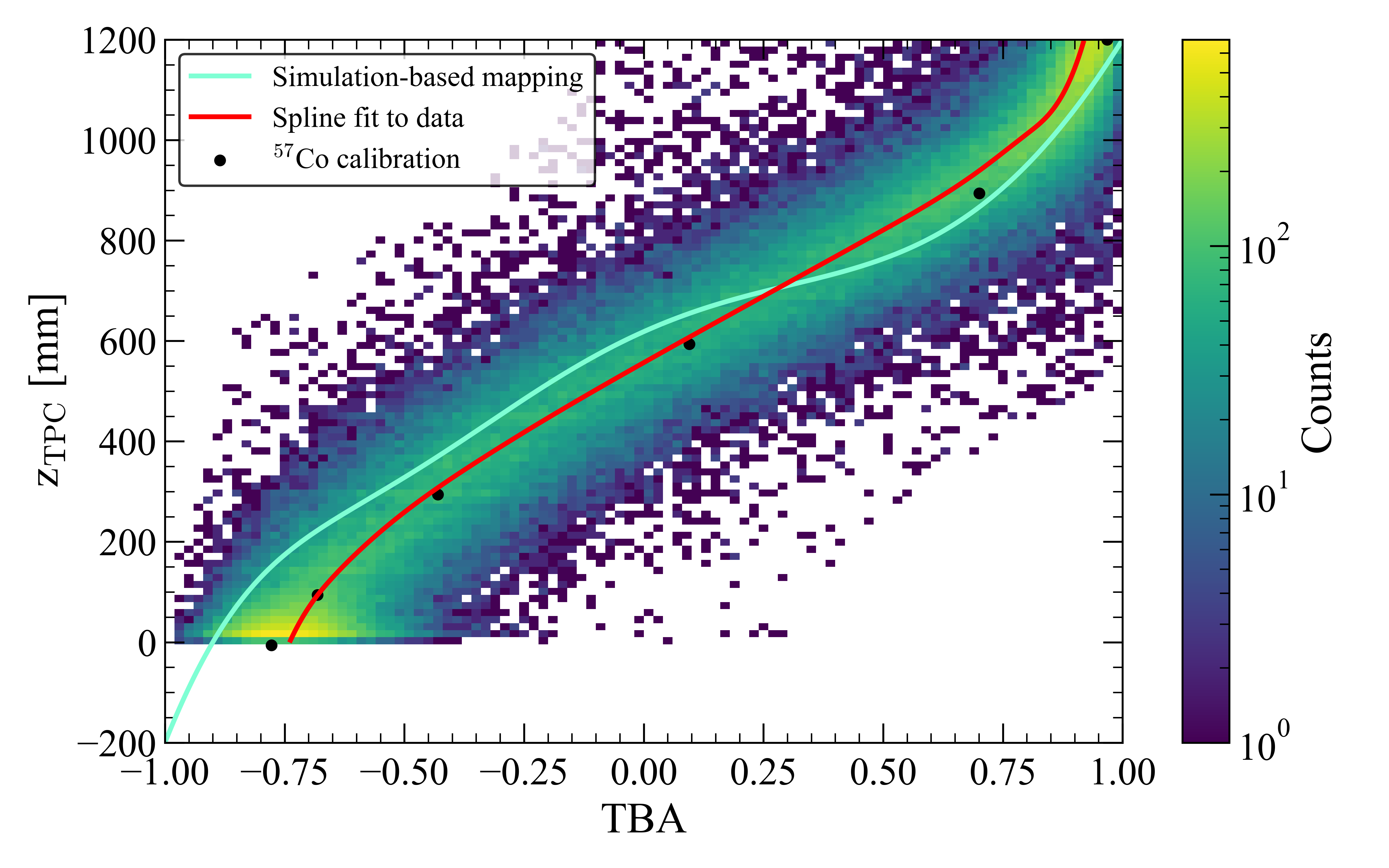}
	\caption{Comparison of data-driven and simulation-based TBA-to-$z$ mappings in the Skin Barrel region, where the TBA is calculated including the Dome PMTs. The color map shows the distribution of correlated Skin–TPC pulse pairs from internal gamma-ray sources in the WS2022 dataset, plotted in the (TBA, $z_{\mathrm{TPC}}$) plane. The green curve shows the mapping derived from detector simulations, while the red curve is a five-node spline fit to the data. Black points indicate measured TBA values from monoenergetic \textsuperscript{57}Co gamma-ray calibration sources at known vertical positions, with TBA uncertainties smaller than the marker size. The data-driven spline better aligns with the calibration points and reduces bias compared to the simulation-derived model, particularly at low $z$. The $z_{\mathrm{TPC}}$ position cannot be defined below 0 (the location of the cathode). }
	\label{fig:TBA to z}
\end{figure}

To produce a robust mapping of TBA-to-$z$, we leverage the subset of internal backgrounds producing multiphoton events capable of depositing energy both in the outer TPC region and in the adjacent Skin. The vertical position of the TPC scatter is reconstructed via drift time with mm-scale accuracy, while the accompanying Skin signal defines the corresponding TBA value. A large, high-purity sample of (TBA, z) pairs can be extracted directly from WS2022 data by requiring geometric alignment and large energy deposits to suppress accidental overlaps.

The resulting distribution reveals a smooth but nonlinear dependence of TBA on vertical position, which is well captured by a constrained cubic spline interpolation (Fig.~\ref{fig:TBA to z}). Compared to early simulation-derived mapping, the data-driven approach provides improved consistency with mono-energetic gamma-ray source deployments, including $^{57}$Co, across the full extent of the instrumented Skin. The calibration exhibits substantial corrections in the lower Barrel region, where the optical path lengths and PMT angular response deviate most strongly from the assumptions used in simulation.

\subsection{Gamma Energy Scale ($E_\gamma$)}

\begin{figure}
	\centering
	\begin{tikzpicture}
		\node[anchor=south west, inner sep=0] (img1) at (0,0) 
			{\includegraphics[width=0.49\linewidth]{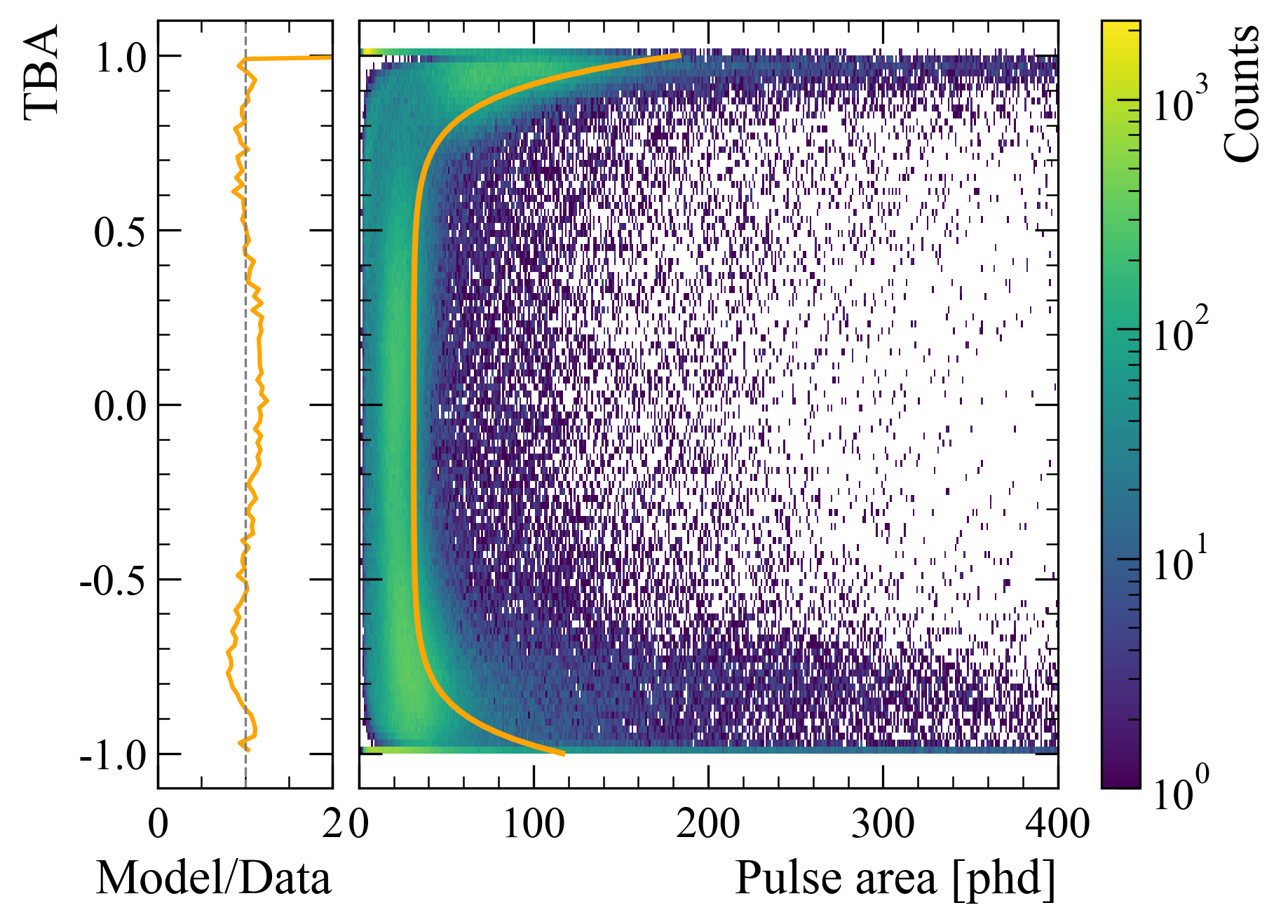}};
		\node[anchor=north west, fill=white, inner sep=2pt, font=\small, yshift=10pt] at (img1.north west) {\textbf{(a)}};
	\end{tikzpicture}
	\hfill
	\begin{tikzpicture}
		\node[anchor=south west, inner sep=0] (img2) at (0,0) 
			{\includegraphics[width=0.49\linewidth]{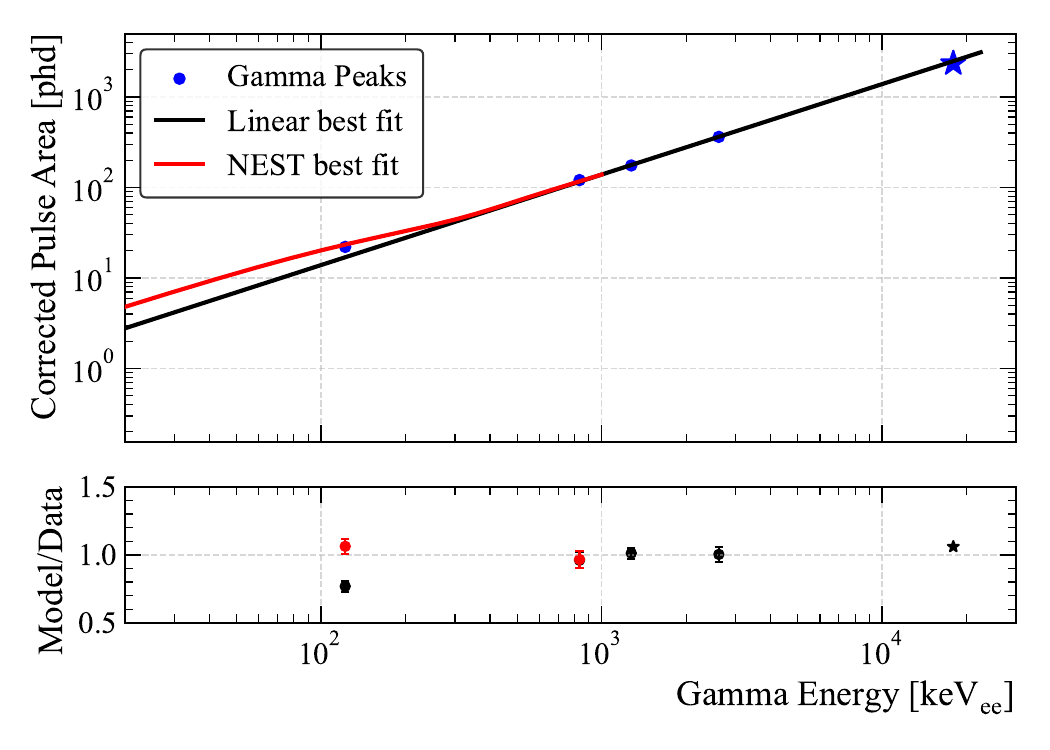}};
		\node[anchor=north west, fill=white, inner sep=2pt, font=\small, yshift=10pt] at (img2.north west) {\textbf{(b)}};
	\end{tikzpicture}
	\caption{Gamma energy calibration in the Skin Barrel. \textbf{(a)} Distribution of TBA versus S1 pulse area from a $^{57}$Co source deployed through the CSD tube at 240$^\circ$. The orange curve indicates the best-fit hyperbolic cosine model used to describe the vertical dependence of light collection, which is applied to correct the energy response across TBA to be consistent with measurements taken at z = 500 mm (TBA = $-$0.1). The left subpanel shows the ratio of the mean pulse area to the best-fit model as a function of TBA. \textbf{(b)} Corrected S1 pulse areas for $^{57}$Co, $^{54}$Mn, $^{22}$Na, and $^{228}$Th sources as a function of known gamma energy. The red and black curves represent best-fit NEST-based and linear models, respectively. The error bars represent the standard deviations for mean area fitted at different source positions. The blue star indicates the $^{222}$Rn peak at its equivalent energy of 17.9 MeV$_\text{ee}$ (Sec. \ref{sec:Alphas}).}
    \label{fig:Skin gamma}
\end{figure}

The gamma-ray energy response of the Skin Barrel was characterized using the monoenergetic source deployments described above. To account for vertical positional dependence, a TBA-based correction map was applied. For each source and deployment height, the pulse area distribution was binned in TBA, and a hyperbolic cosine model:
\[f(\text{TBA}) = \alpha\cosh(\beta \text{TBA}+\delta)+\gamma,\]
where $\alpha$, $\beta$, $\gamma$, and $\delta$ are free parameters fitted to $^{57}$Co calibration data taken on a CSD tube at 240$^\circ$, was used to describe the vertical dependence. This model provided a stable and accurate fit across the full TBA range, as shown in Fig.~\ref{fig:Skin gamma}(a).

With the position dependence mitigated, a mapping between corrected pulse area and deposited energy can be constructed using the Noble Element Simulation Technique (NEST)~\cite{szydagis2025review}. Fig.~\ref{fig:Skin gamma}(b) shows the NESTv2.3.12-based model of pulse area versus deposited energy assuming an average electric field of 5 kV/cm for gamma energies below 1 MeV and a linear fit at higher energies. The energy scale was normalized using the 122 keV peak from $^{57}$Co, which produced an S1 pulse of approximately 21 phd at the center of the Skin Barrel, yielding a conversion factor of 5.8 keV/phd. No significant azimuthal dependence was observed across the three CSD tubes, confirming the largely symmetric optical response in the Skin.

The calibrations shown in Fig.~\ref{fig:Skin gamma} are from a post-WS2022 calibration campaign, and comprehensive CSD calibrations are not yet available for WS2024. The existing map is applicable to WS2024 with an estimated 5--10\% systematic increase in light yield due to the changed electric field, based on $^{228}$Th data taken in the WS2024 conditions. An alternative position correction map derived from alphas is described in the next section, with alphas offering a more field-independent response and improved position resolution given their higher signal size. The difference between the two maps is taken as a systematic uncertainty on the veto energy threshold discussed in Section~\ref{sec:Veto}.




The CSD gamma calibrations do not efficiently illuminate the Dome region because of the CSD geometry, so energy calibration in the Dome using gamma sources is not feasible. Consequently, the energy scale determination in the Dome relies on an alternative method using alpha decays, discussed in Sec.~\ref{sec:PDVeto}.

\section{Reconstructing Alphas in the Skin}\label{sec:Alphas}

The decay chains from radioactive Rn isotopes produce multiple high-energy alpha particles that serve as a complementary calibration source throughout the science runs. These alphas are easily identified due to their large scintillation signals and are commonly used to assess the energy and position reconstruction performance of the detector. In particular, the following decays from $^{222}$Rn:
\[
^{222}_{86}\mathrm{Rn}~(3.8~\mathrm{d}) \longrightarrow {}^{218}_{84}\mathrm{Po}~(Q_\alpha = 5.6~\mathrm{MeV})
\]
\[
^{218}_{84}\mathrm{Po}~(3.1~\mathrm{m}) \longrightarrow {}^{214}_{82}\mathrm{Pb}~(Q_\alpha = 6.1~\mathrm{MeV})
\]
\[\vdots\]
\[
^{214}_{84}\mathrm{Po}~(164~\si{\micro \second}) \longrightarrow {}^{210}_{82}\mathrm{Pb}~(Q_\alpha = 7.8~\mathrm{MeV});
\]
and from $^{220}$Rn:
\[
^{220}_{86}\mathrm{Rn}~(55.6~\mathrm{s}) \longrightarrow {}^{216}_{84}\mathrm{Po}~(Q_\alpha = 6.4~\mathrm{MeV})
\]
\[
^{216}_{84}\mathrm{Po}~(0.15~\mathrm{s}) \longrightarrow {}^{212}_{82}\mathrm{Pb}~(Q_\alpha = 6.9~\mathrm{MeV})
\]
\[\vdots\]
\[
^{212}_{84}\mathrm{Po}~(0.299~\si{\micro \second}) \longrightarrow {}^{208}_{82}\mathrm{Pb}~(Q_\alpha = 9.0~\mathrm{MeV});
\]
provide a clean sample to study detector resolution and to identify bulk liquid xenon flow. The short half-lives of the Rn daughters enable time-correlated tagging of decay pairs, which aids in the identification and selection of alpha events.

\begin{figure}
	\centering
    \begin{tikzpicture}
		\node[anchor=south west, inner sep=0] (img1) at (0,0) 
			{\includegraphics[width=0.9\linewidth]{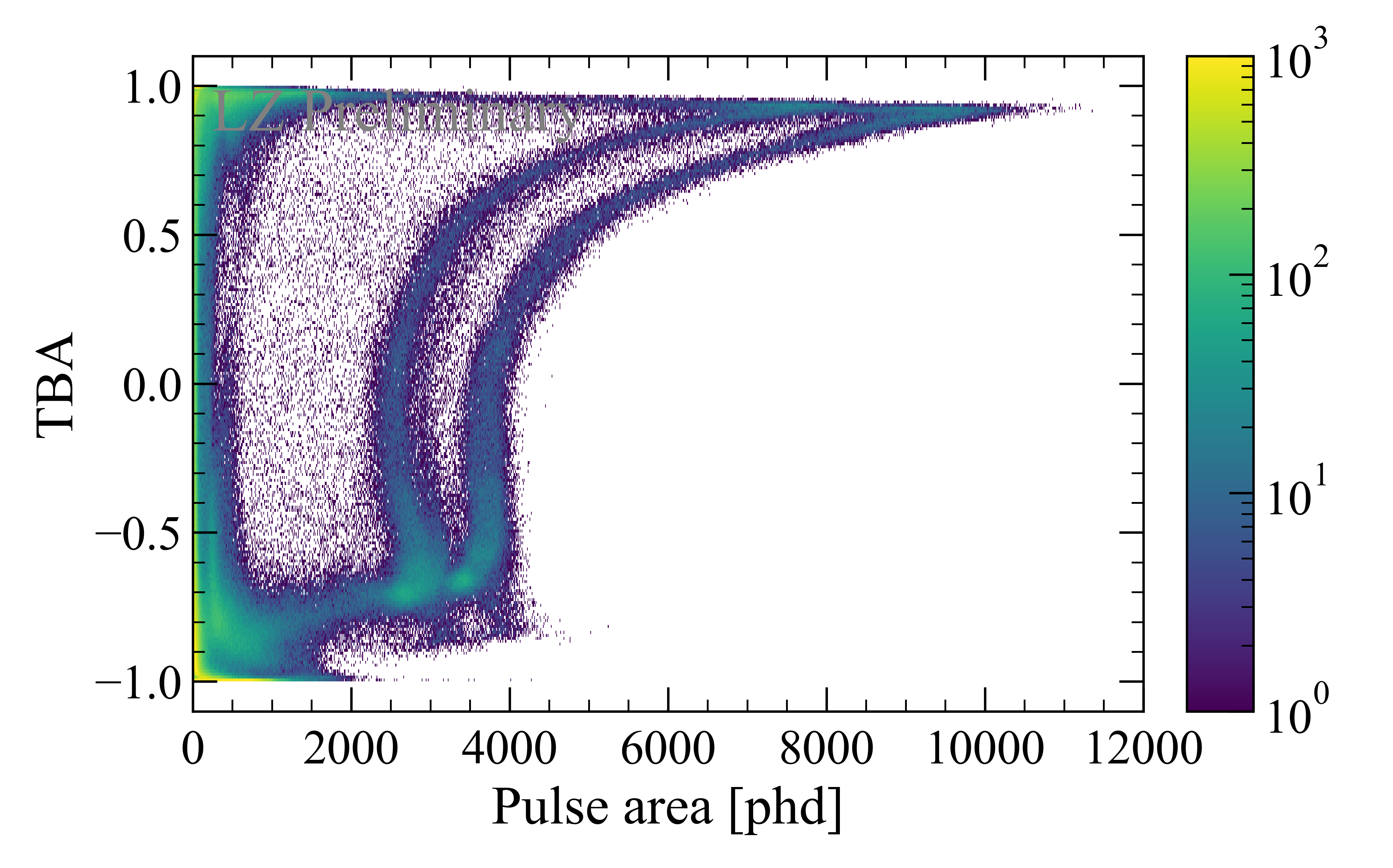}};
		\node[anchor=north west, fill=white, inner sep=2pt, font=\small] at (img1.north west) {\textbf{(a)}};
	\end{tikzpicture}
	\hfill
	\begin{tikzpicture}
		\node[anchor=south west, inner sep=0] (img2) at (0,0) 
			{\includegraphics[width=0.9\linewidth]{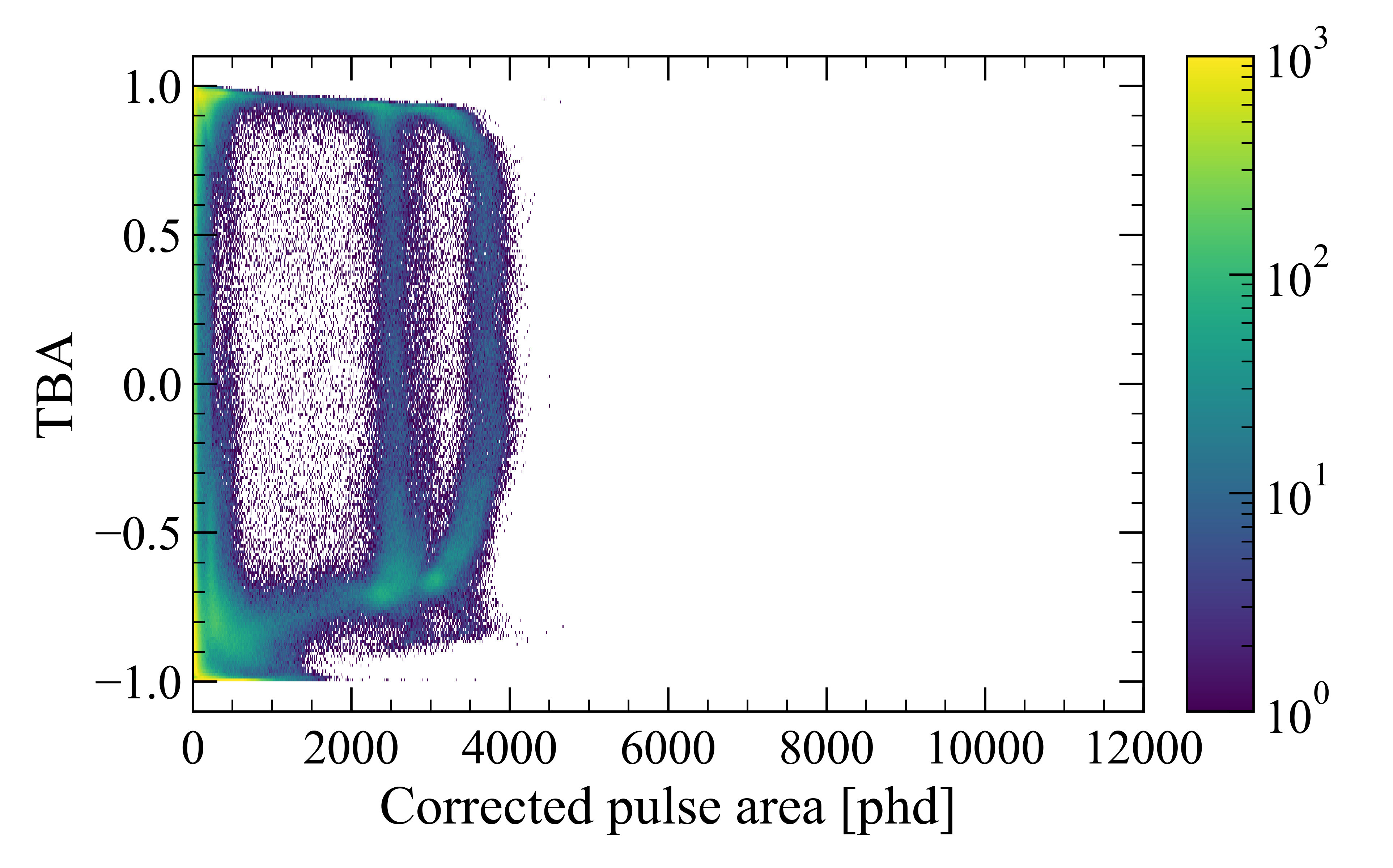}};
		\node[anchor=north west, fill=white, inner sep=2pt, font=\small] at (img2.north west) {\textbf{(b)}};
	\end{tikzpicture}
    \caption{Two-dimensional distribution of TBA versus pulse area for events during a dedicated $^{220}$Rn calibration run in November 2022. The color scale represents event counts on a logarithmic scale. Two distinct bands are visible, corresponding to the combined $^{220}$Rn/$^{216}$Po alpha peaks and the higher-energy $^{212}$Po peak. \textbf{(a)} Before correction, the curvature of each band reflects position-dependent variation in light collection efficiency, with events near the top PMTs producing larger pulse areas. \textbf{(b)} After applying position-dependent corrections based on the $^{220}$Rn/$^{216}$Po, the dependence is largely mitigated.}
	\label{fig:rn220_hist2d}
\end{figure}

\begin{figure}
    \centering
    \includegraphics[width=0.9\linewidth]{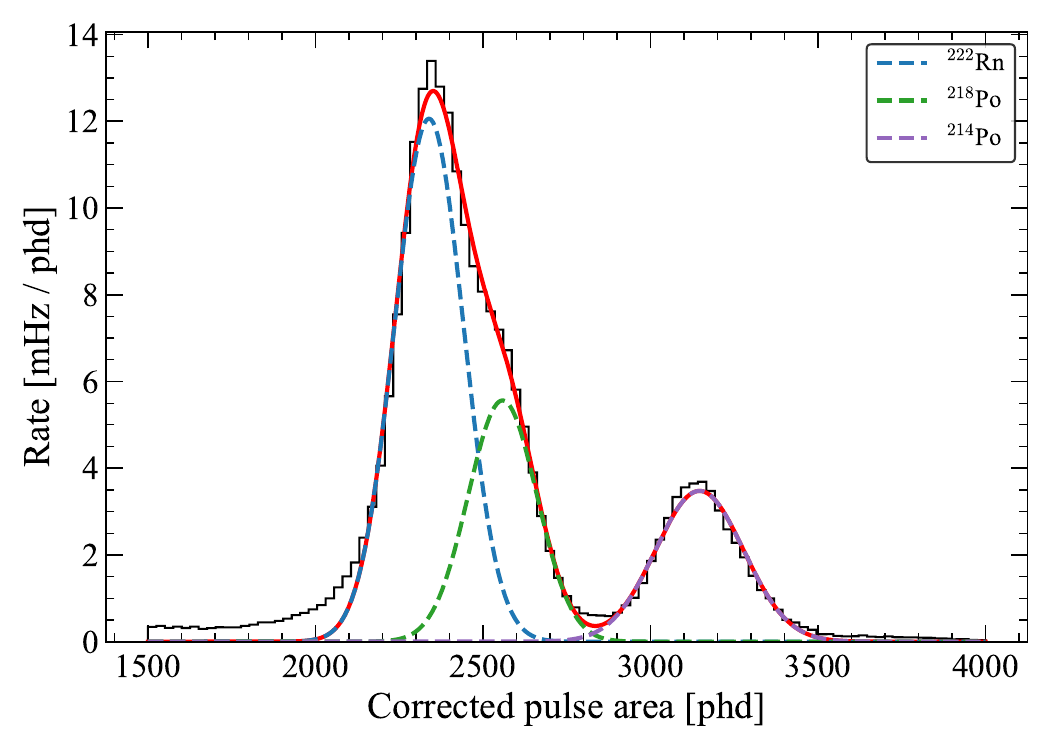}
    \caption{Distribution of corrected pulse areas from alpha decays in the Skin Detector during WS2024. The black histogram shows the measured spectrum, with three fitted Gaussian peaks corresponding to $^{222}$Rn (blue dashed), $^{218}$Po (green dashed), and $^{214}$Po (purple dashed). The combined fit is shown in red.}
    \label{fig:alpha_peakfit}
\end{figure}

Position corrections derived from dedicated $^{220}$Rn calibration runs are applied to the alpha data, shown in Fig.~\ref{fig:rn220_hist2d}. After correction, the alpha peaks from $^{222}$Rn, $^{218}$Po, and $^{214}$Po in the central region of the Skin ($-$0.5 < TBA < 0.5) are fitted to determine the decay rates for each isotope (Fig.~\ref{fig:alpha_peakfit} and Table~\ref{tab:alpha_rates}). The measured activity of $^{222}$Rn is about 50\% higher in the Skin compared to the TPC for both WS2022~\cite{aalbers2023background} and WS2024~\cite{aalbers2025flow}, possibly supporting the hypothesis that the dominant source of radon is the cryostat~\cite{aalbers2023background}. With the addition of the Skin trigger in WS2024, the activity of $^{220}$Rn-$^{216}$Po could be determined separately by time-correlated pairfinding as described in~\cite{aalbers2025flow}. $^{220}$Rn activity during WS2022 was assumed to be the same.

While the $^{222}$Rn and $^{218}$Po peaks are not fully separable, the fit indicates that the $^{218}$Po rate is a bit less than half that of $^{222}$Rn, with $^{214}$Po showing a similar but smaller reduction. Studies in the TPC observed that $\sim 50$\% of alpha decay daughters are charged~\cite{aalbers2025flow}. Charged daughters in the Skin would rapidly drift radially and be captured by the inner wall, providing a potential explanation for the observed suppression, but a detailed study in the Skin has not been performed.

\begin{table}[b]
\caption{Table of measured alpha activities in the Skin. The $^{222}$Rn, $^{218}$Po, and $^{214}$Po rates are measured by fitting Gaussian peaks to the corrected pulse area spectrum (Fig.~\ref{fig:alpha_peakfit}). The $^{220}$Rn/$^{216}$Po rate is measured by time-correlated pairfinding, which was not available for WS2022 due to the absence of the Skin trigger.}
\vspace{6pt}
\centering
\begin{tabular}{|c|c|c|}
\hline
Radon Isotope & WS2022 Activity [µBq/kg] & WS2024 Activity [µBq/kg] \\
\hline
$^{222}$Rn & 7.27 $\pm$ 0.61 & 6.70 $\pm$ 0.28 \\
$^{218}$Po & 3.05 $\pm$ 0.56 & 3.00 $\pm$ 0.29 \\
$^{214}$Po & 3.08 $\pm$ 0.44 & 2.31 $\pm$ 0.22 \\
$^{220}$Rn/$^{216}$Po & - & 0.164 $\pm$ 0.003 \\
\hline
\end{tabular}
\label{tab:alpha_rates}
\end{table}

\subsection{Position Resolution}

\begin{figure}
	\centering
	\begin{tikzpicture}
		\node[anchor=south west, inner sep=0] (img1) at (0,0) 
			{\includegraphics[width=0.49\linewidth]{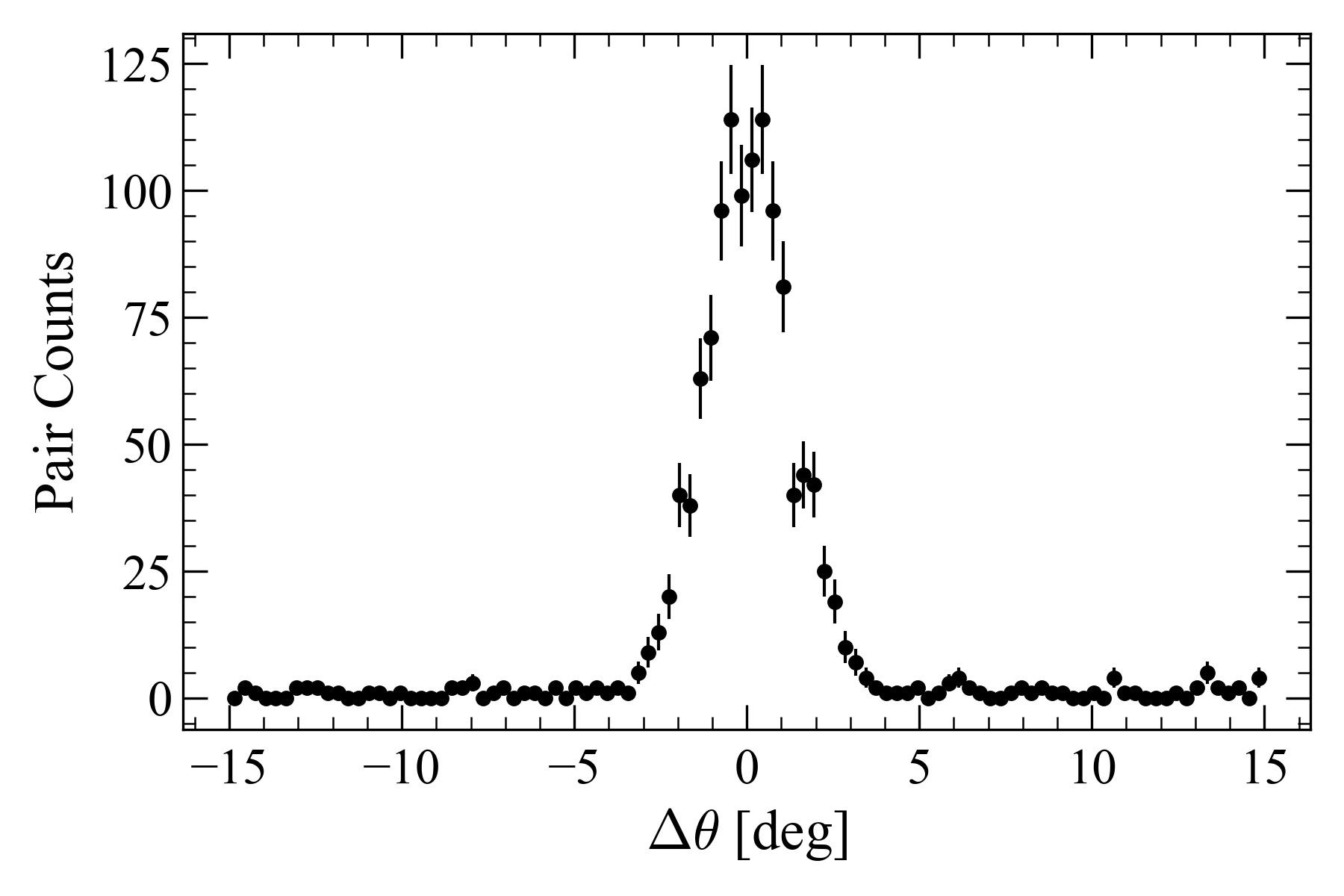}};
		\node[anchor=north west, fill=white, inner sep=2pt, font=\small] at (img1.north west) {\textbf{(a)}};
	\end{tikzpicture}
	\hfill
	\begin{tikzpicture}
		\node[anchor=south west, inner sep=0] (img2) at (0,0) 
			{ \includegraphics[width=0.49\linewidth]{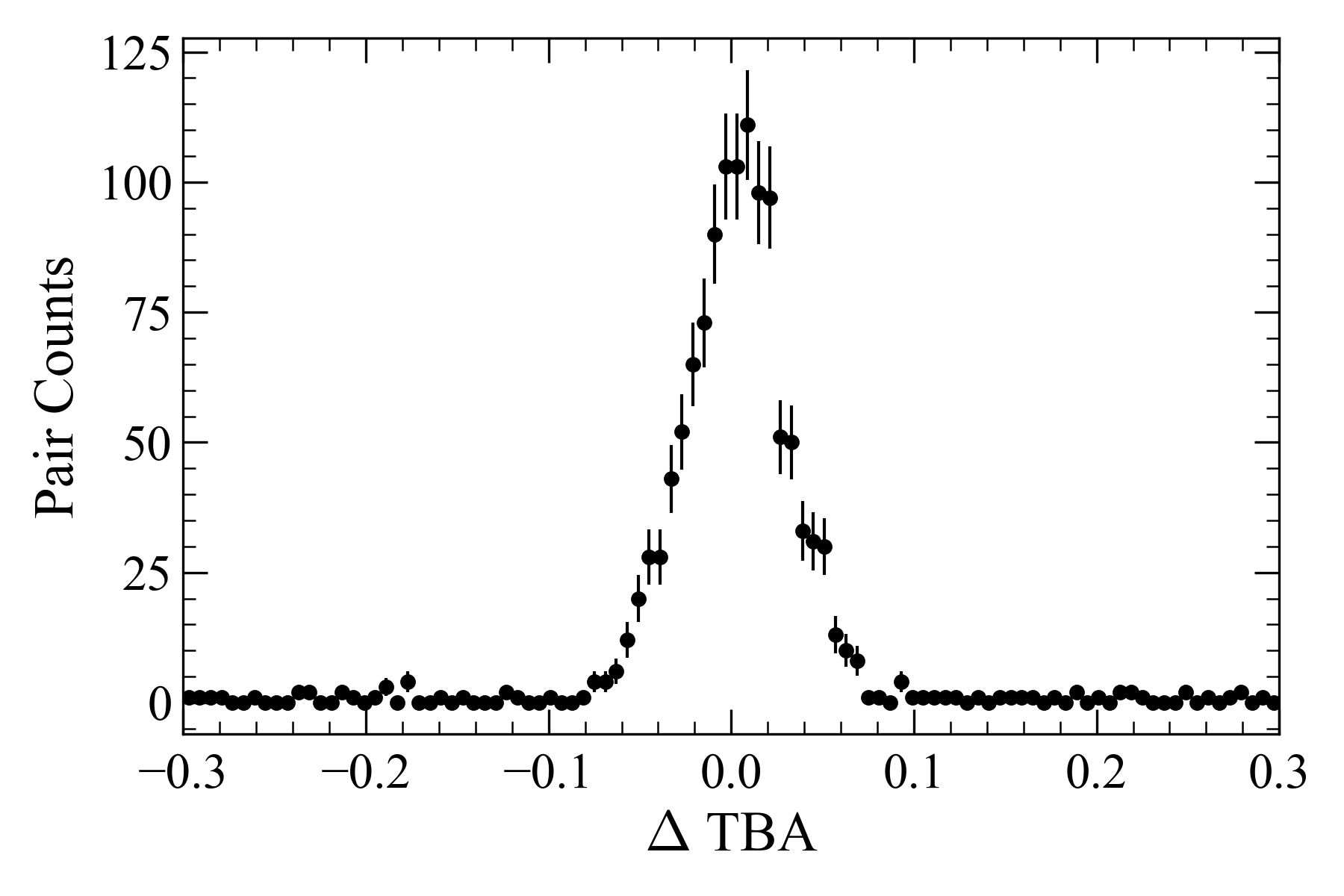}};
		\node[anchor=north west, fill=white, inner sep=2pt, font=\small] at (img2.north west) {\textbf{(b)}};
	\end{tikzpicture}
	\caption{Reconstructed spatial differences between $\mathrm{^{220}Rn}$–$\mathrm{^{216}Po}$ pairs, used to assess the position resolution of the Skin Detector. 
    \textbf{(a)} Azimuthal angle difference $\Delta\theta$, exhibiting a standard deviation of approximately $1.2^\circ$, corresponding to a lateral resolution of $\sim$15~mm. 
    \textbf{(b)} Difference in top-bottom asymmetry $\Delta\mathrm{TBA}$, serving as a proxy for vertical displacement. The standard deviation is approximately $0.026$, corresponding to a vertical resolution of $\sim$13~mm. }
	\label{fig:Position}
\end{figure}

To quantitatively determine the Skin’s position resolution, we exploit the coincidence between $^{220}$Rn and $^{216}$Po alpha decays, as these decays can be assumed to occur at the same physical location given the 146 ms half-life of $^{216}$Po. From calibration runs with injected $^{220}$Rn during November 2022, the decay pairs are selected by requiring time separations less than 1 second with corrected pulse areas between 2000 and 2850 phd.

With this dataset, the angular resolution is evaluated by computing the difference in reconstructed $\theta$ between the paired decays (Fig.~\ref{fig:Position}). The resulting distribution is centered at zero, with a standard deviation of 1.2~degrees, corresponding to approximately 15~mm in azimuthal position. Similarly, the TBA separation has a standard deviation of 0.026 corresponding to approximately 13~mm in vertical position, using the positional mapping based on gamma calibrations. The distribution is centered at zero, confirming that the reconstruction is unbiased and that the selected pairs are spatially coincident. Note that these values correspond to the pairwise resolution, and the spatial resolutions for single events would be scaled by 1/$\sqrt{2}$.

These measurements demonstrate that the Skin Detector has sufficient spatial resolution to localize energy depositions within tens of millimeters in both $\theta$ and $z$ for large enough signals that are not limited by Poisson uncertainties. While full three-dimensional reconstruction is not possible due to the lack of radial sensitivity, the available information enables effective spatial tagging of background events and supports the Skin’s role in vetoing and background modeling.

\subsection{Liquid Flow in the Skin}

\begin{figure}
	\centering
    \begin{tikzpicture}
		\node[anchor=south west, inner sep=0] (img1) at (0,0) 
			{\includegraphics[width=0.9\linewidth]{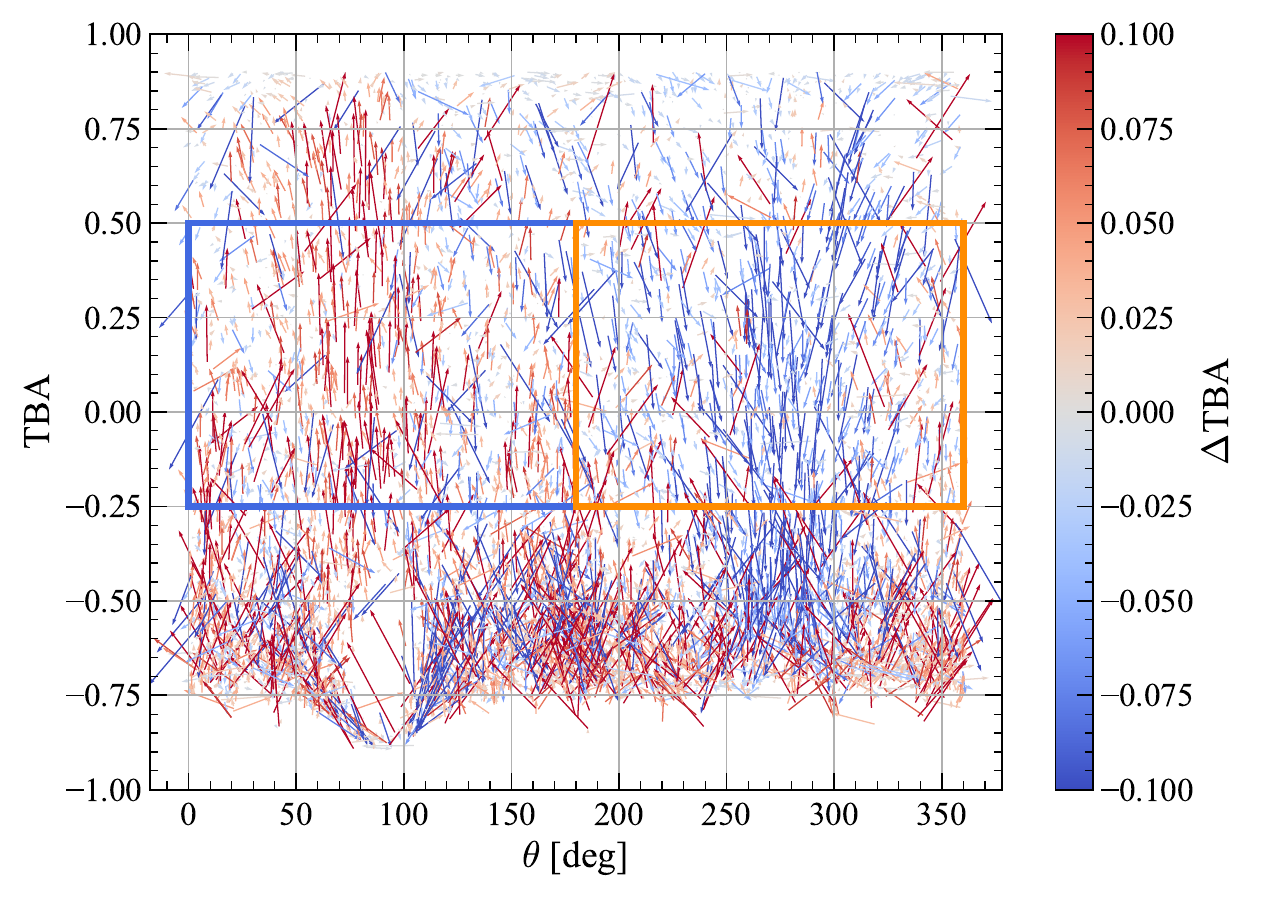}};
		\node[anchor=north west, fill=white, inner sep=2pt, font=\small] at (img1.north west) {\textbf{(a)}};
	\end{tikzpicture}
	\hfill
	\begin{tikzpicture}
		\node[anchor=south west, inner sep=0] (img2) at (0,0) 
			{\includegraphics[width=0.9\linewidth]{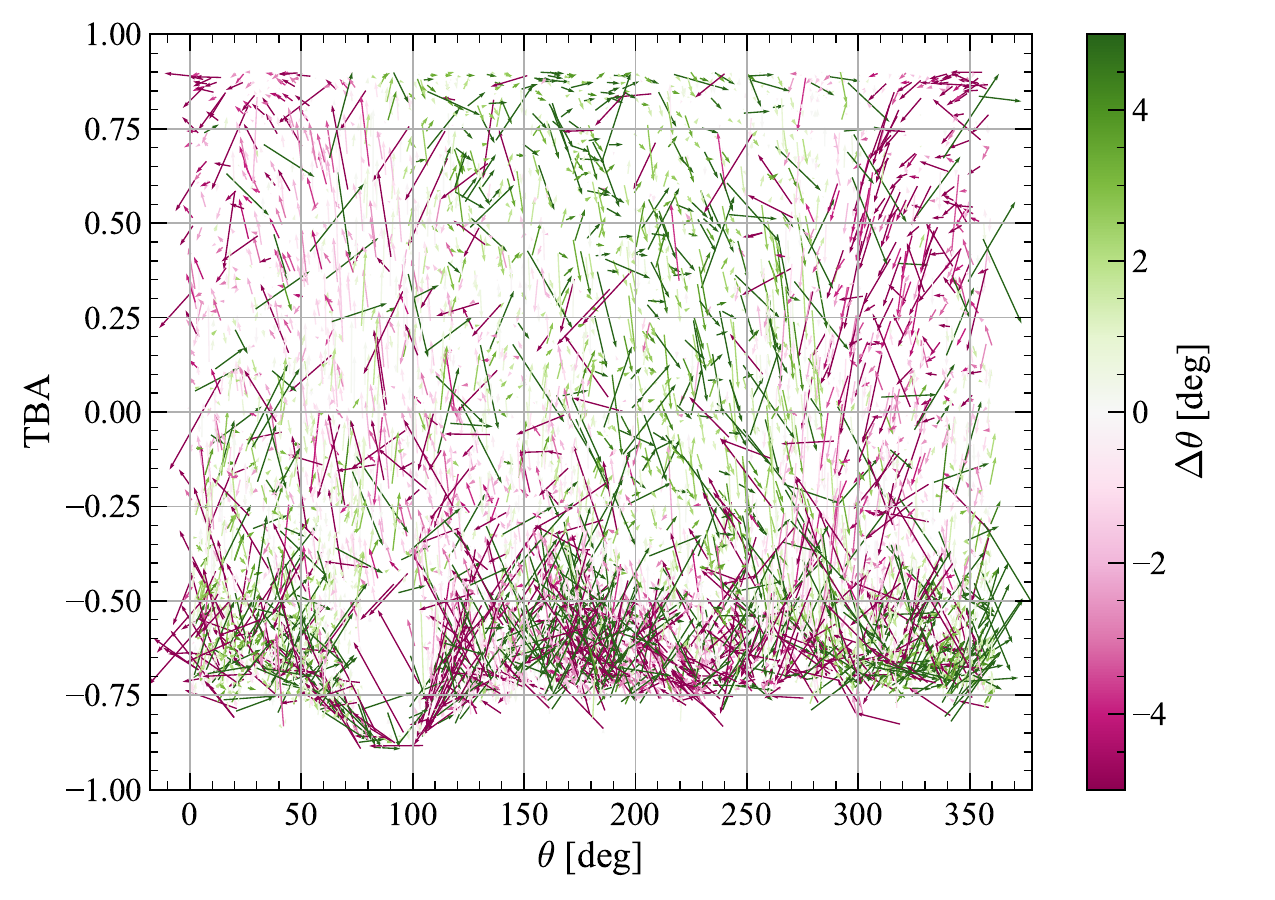}};
		\node[anchor=north west, fill=white, inner sep=2pt, font=\small] at (img2.north west) {\textbf{(b)}};
	\end{tikzpicture}
	\caption{Visualizations of xenon flow in the Skin derived from $^{222}$Rn–$^{218}$Po alpha pairs with time separation less than 10 seconds. The arrows indicate the reconstructed vertices of the decay pairs. The same flow field is shown twice, with different colors to highlight distinct components of the motion. The HV conduit is located near the bottom at $\theta = 90^\circ$, partly impacting reconstruction.
    \textbf{(a)} Vertical component of the flow field, color-coded by $\Delta\mathrm{TBA}$. Velocity distributions in the blue and orange boxes are shown in Fig.~\ref{fig:RnFlowHist}.
    \textbf{(b)} Horizontal component, color-coded by $\Delta\theta$.}
    \label{fig:RnFlow}
\end{figure}

\begin{figure}
	\centering
    \includegraphics[width=0.9\linewidth]{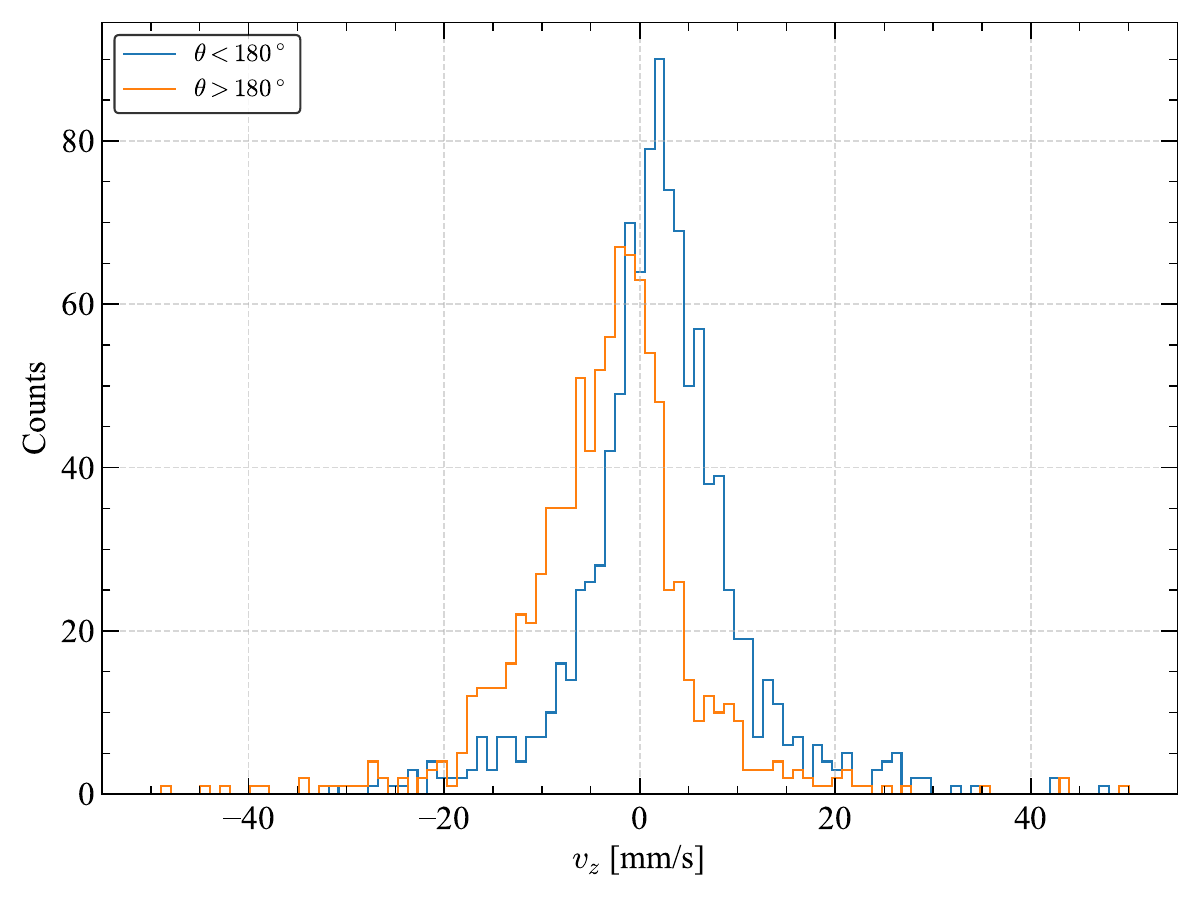}
	\caption{Distribution of vertical velocities $v_z$ for $\theta < 180^\circ$ and $\theta > 180^\circ$ regions near the center of the detector ($-$0.25 < TBA < 0.5), showing asymmetry indicative of bulk motion. The average flow velocities for each region are 2.0 mm/s and $-$3.8 mm/s, respectively.}
    \label{fig:RnFlowHist}
\end{figure}

In~\cite{aalbers2025flow}, distinct flow states in the TPC were categorized as "isolated" or  "mixed" depending on the thermodynamics of the circulation system, with liquid flow measurable in the less-turbulent isolated state by examining Rn decay pairs. 
After establishing the spatial resolution of the Skin, the liquid flow in the Skin can also be studied with decay pairs involving $^{222}$Rn and $^{218}$Po, which are separated by a multi-minute half life. Since $^{222}$Rn has a much higher concentration in the detector during regular operations compared to $^{220}$Rn, the events were selected from the isolated state of WS2024 by requiring a time separation between 1 and 10 seconds, similar pulse area requirements to $^{220}$Rn, and maximum spatial displacements of $\Delta$ TBA $<0.25$ and $\Delta \theta < 25$ degrees to reduce accidental pairings. Finding pairs with longer time separations is challenging as liquid flow and drift leads pairs to drift out of the allowed spatial cuts. 

To visualize this motion, the displacement vectors between selected $^{222}$Rn-$^{218}$Po pairs are plotted in $\theta$–TBA space and colored by the vertical and azimuthal components of the motion. Figure~\ref{fig:RnFlow}(a) displays $\Delta\mathrm{TBA}$, which is proportional to vertical displacement and reveals convective flow patterns. A net upward trend is observed in certain regions, towards an azimuth angle of $90^\circ$, while a symmetric downward trend is observed near $270^\circ$. Figure~\ref{fig:RnFlow}(b) presents the azimuthal component $\Delta\theta$, which shows a similar structure with horizontal circulation.

To further quantify the vertical motion, we convert $\Delta\mathrm{TBA}$ to physical velocity estimates using the empirically determined TBA-to-$z$ mapping and the known decay time between the RnPo pairs. The resulting $v_z$ distribution is shown in Fig.~\ref{fig:RnFlowHist} for two different $\theta$ regions. A clear difference in the peak position is observed, where the two regions exhibit vertical flows at average rates of 2.0 mm/s and $-$3.8 mm/s, respectively. This flow rate corresponds to the LXe fully circulating within the Skin volume approximately within 10 minutes, and is significantly faster than the average flow rate observed in the TPC of $<$ 0.1 mm/s in isolated states~\cite{aalbers2025flow}. The flow patterns remain stable throughout the data-taking period from August 2023 to May 2025, supporting the existence of stable flow cells within the Skin. 

\section{Skin Veto Performance}\label{sec:Veto}

\begin{figure}
	\centering
	\includegraphics[width=0.98\linewidth]{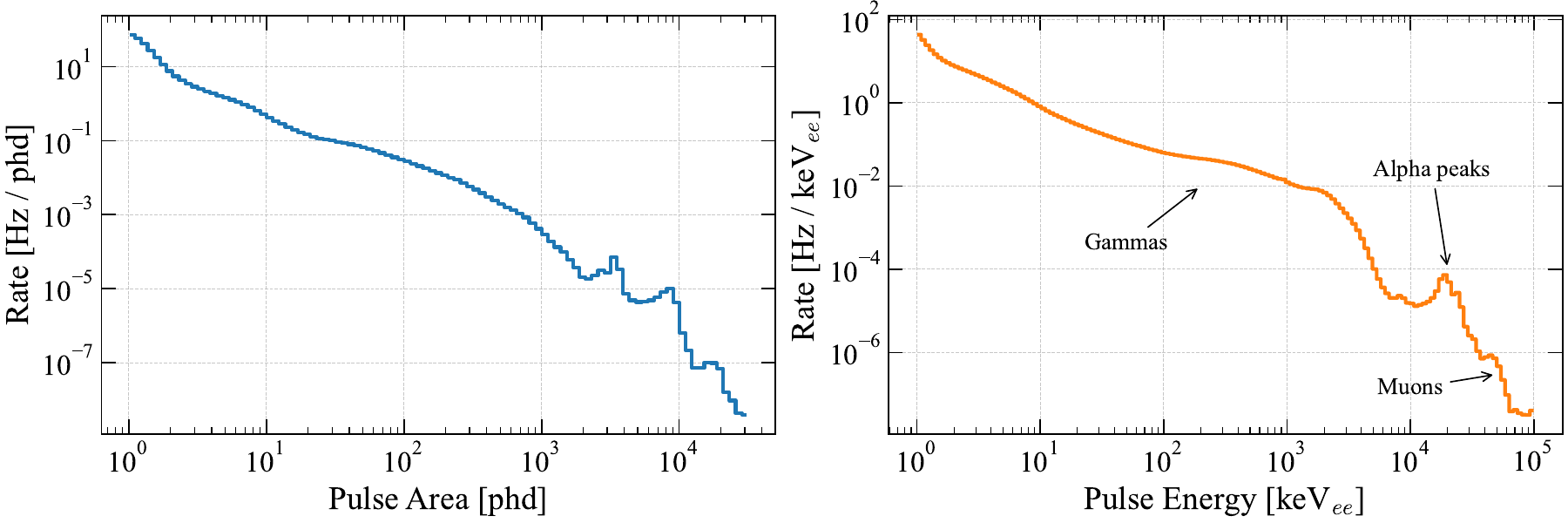}
	\caption{Pulse area and energy distribution for Skin pulses acquired in the post-trigger window of all trigger types during February 2025. The low-area population is dominated by gamma-induced events, the two intermediate-area peaks correspond to alpha decays, and the largest areas from muons. The total rate of all pulses in the Skin is 52 Hz.}
    \label{fig:rate}
\end{figure}

Throughout WS2022 and WS2024, the Skin maintained stable operating conditions with the exception of spurious light emissions near the cathode between the two runs~\cite{aalbers2025dark}. Figure~\ref{fig:rate} shows a snapshot of the Skin activity during February 2025, illustrating a typical period of detector operation. The total rate of all signals classified as pulses by the analysis is 52 Hz, measured in unbiased windows from randomly triggered events and including signals down to single photoelectrons. The low-area population, dominated by gamma-induced interactions, represents the majority of detected events, while intermediate-area peaks correspond to alpha decays from radon progeny. The highest pulse areas arise from cosmic-ray muons traversing the Skin Detector volume (once or twice).


\subsection{Prompt and Delayed Veto}
\label{sec:PDVeto}

The Skin veto logic is divided into two categories: prompt vetoes, which target fast interactions coincident with the primary TPC signal, and delayed vetoes, which identify subsequent energy depositions from neutron capture gamma cascades. The specific conditions for each veto type are summarized in Table~\ref{tab:veto_conditions}, where the time window is calculated relative to the TPC S1 signal. The veto thresholds are chosen such that the deadtime from false positives induced by noise and dark counts is not significant.

\begin{figure}
	\centering
	\includegraphics[width=0.9\linewidth]{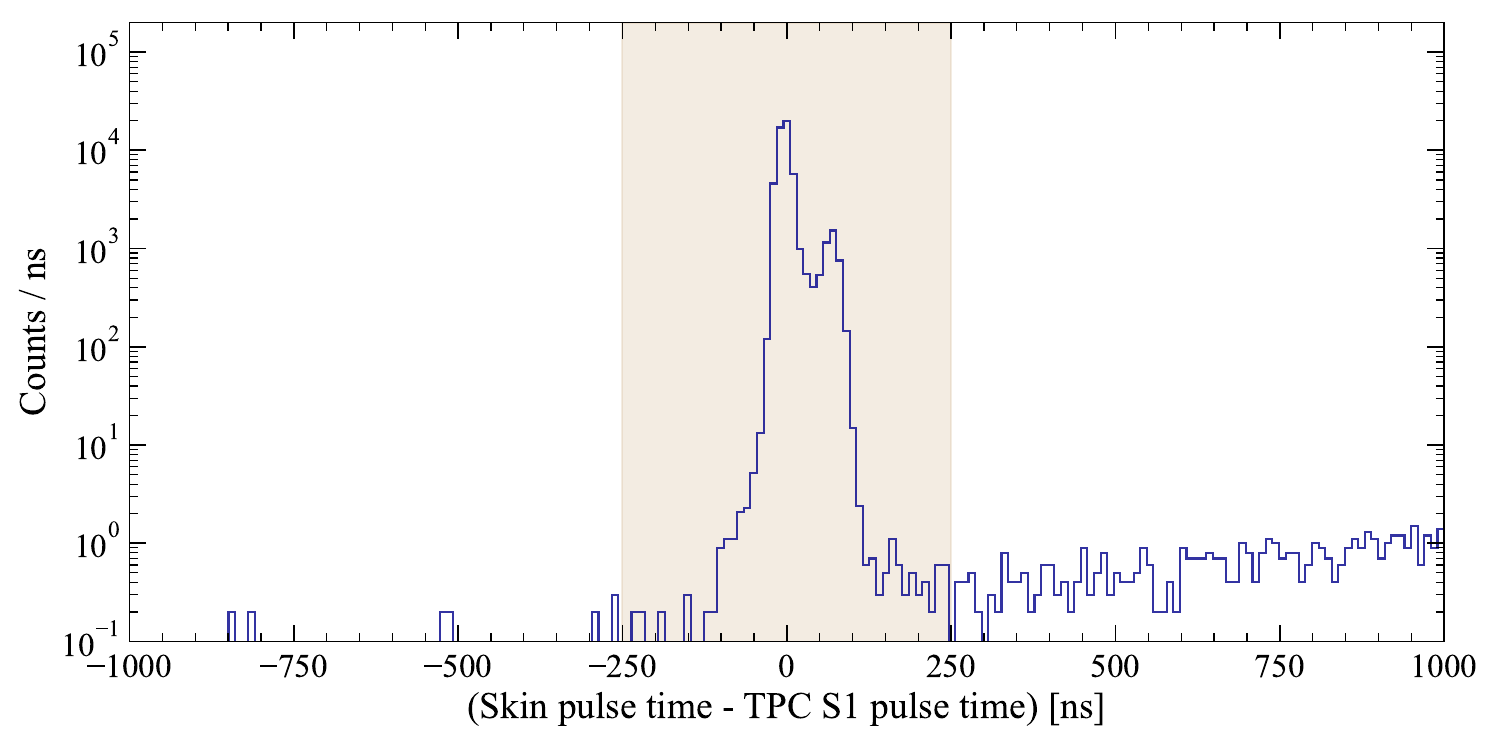}
	\caption{Distribution of time differences between TPC S1 and Skin pulses above the prompt veto threshold, for single-scatter events acquired in February 2025. The shaded region ($\pm$250 ns) indicates the coincidence window used to define the Skin prompt veto in WS2024.}
    \label{fig:skin_timing}
\end{figure}

The prompt veto is designed to reject events involving fast gamma rays or neutrons that interact nearly simultaneously in the TPC and Skin. Figure~\ref{fig:skin_timing} shows the time difference of the start of the S1 pulse in the TPC and Skin pulses for single-scatter events. The small peak at positive time differences comes exclusively from pulses in the Dome PMTs, illustrating a timing misalignment that will be corrected in future analyses. The residual tail extending beyond 250 ns may suggest contributions from light leakage from the TPC into the Skin volume or fluorescent backgrounds following large light pulses, and are a subject for future study.

\begin{table}[t]
    \caption{Summary of prompt and delayed veto conditions for WS2022 and WS2024. The pulse rates that satisfy the pulse area and coincidence conditions are measured from randomly triggered data in February 2025.}
    \vspace{6pt}
    \centering
    \small
    \begin{tabular}{|c|c|c|c|c|}
    \hline
        & Pulse Area [phd] & PMT Coincidence & Time Window [\si{\micro\second}] & Pulse Rate [Hz] \\
        \hline
        WS2022 Prompt  & $>$2.5 & $\geq$3 & $\pm$0.5 & 10.6 \\
        WS2022 Delayed & $>$50  & $\geq$56 & 0.5 -- 1200 & 0.3 \\
        WS2024 Prompt  & $>$2.5 & $\geq$3 & $\pm$0.25 & 10.6 \\
        WS2024 Delayed & $>$46  & $\geq$3 & 0.25 -- 600 & 6.3 \\
        \hline
    \end{tabular}
    \label{tab:veto_conditions}
\end{table}

The delayed veto targets radiogenic neutrons that scatter in the TPC and subsequently thermalize and capture on surrounding detector materials, particularly gadolinium in the OD, producing delayed gamma-ray cascades up to several hundred microseconds after the initial S1. While the OD is the primary veto system for neutron captures, the gamma-ray cascades can also deposit energy in the Skin. The delayed veto applies higher pulse area and PMT coincidence requirements over a longer time window following the TPC S1. The WS2022 coincidence requirement was conservatively set to minimize deadtime from accidental coincidences, while the WS2024 thresholds were jointly optimized with the OD delayed veto to improve neutron rejection efficiency while maintaining acceptable deadtime levels.

The calculated prompt and delayed veto thresholds for WS2024 are shown in Table~\ref{tab:veto_threshold}, along with those from alpha-based position corrections. The primary design requirement for the Skin veto system was to achieve a gamma-ray energy threshold of 100~keV electron or gamma equivalent energy (keV$_\text{ee}$) in at least 95\% of the Skin volume, excluding regions with known geometric or optical inefficiencies~\cite{mount2017lux}. The prompt veto threshold is <100 keV$_\text{ee}$ throughout the Skin volume, satisfying the design specifications.  The significantly improved threshold performance in the Barrel relative to the design is attributed to using conservative values for the LXe light yield, PTFE reflectivity, and PTFE coverage in the original simulations.

As discussed earlier, establishing the energy scale in the Dome is challenging due to its complex geometry. The light collection efficiency was determined by scaling the response of $^{220}$Rn using time-correlated pairfinding, where the candidate pulses in the Barrel center ($-$0.5 < TBA < 0.5) had a mean area of 3007$\pm$358~phd, compared to the mean area in the Dome of 439$\pm$219~phd, indicating a relative light collection efficiency to be $\sim$15\%. 

\begin{table}[b]
    \caption{Prompt and delayed veto thresholds for WS2024, corresponding to 2.5 and 46 phd, respectively, with at least 3 PMT coincidence, at representative vertical positions in the Skin Detector. Values are derived from gamma-based position corrections, with the corresponding alpha-based values given in parentheses. Barrel thresholds use CSD gamma calibrations, while the Dome threshold is determined from $^{220}$Rn using time-correlated pairfinding.}
    \vspace{6pt}
    \centering
    \begin{tabular}{|c|c|c|}
    \hline
        Vertical Position [mm] & Prompt Veto Threshold [keV$_\text{ee}$] & Delayed Veto Threshold [keV$_\text{ee}$]\\
        \hline
        1100 & 3.4 (3.8) & 100 (110)\\
        800  & 8.5 (10.7) & 310 (390)\\
        400  & 8.6 (8.9) & 310 (320)\\
        0    & 7.4 (7.0) & 260 (240)\\
        Dome &  85 & 2300\\
        \hline
    \end{tabular}
    \label{tab:veto_threshold}
\end{table}

\subsection{Veto Efficiency}\label{sec:veto_efficiency}

\begin{figure}
	\centering
	\includegraphics[width=0.9\linewidth]{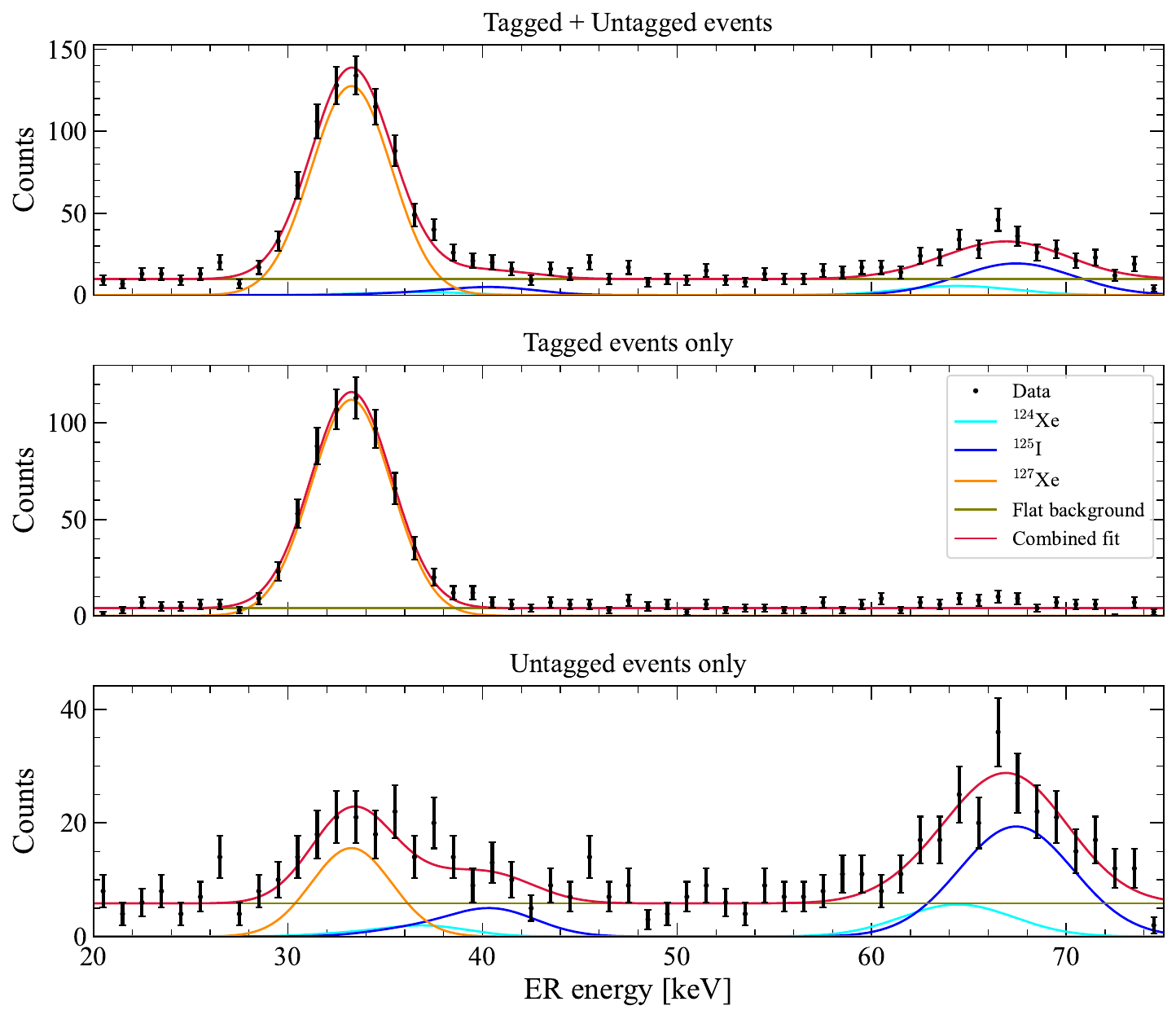}
	\caption{Simultaneous fit to the ER energy spectrum of $^{127}$Xe K-shell events in the TPC, between 3 and 6~cm away from the wall during WS2022. The top panel shows the combined tagged and untagged events, the center panel shows tagged events only, and the bottom panel shows untagged events only. Individual fit components are shown for $^{127}$Xe (orange), $^{125}$I (blue), $^{124}$Xe (cyan), flat background (olive), and total (red).}
    \label{fig:xe127_tagging}
\end{figure}

The prompt veto efficiency for gamma rays can be measured using $^{127}$Xe K-shell electron capture decays which uniformly populate the liquid xenon. The K-shell capture deposits 33.2~keV in X-rays and Auger electrons within the TPC, while the nuclear de-excitation of the resulting $^{127}$I nucleus produces gamma rays (202.9~keV, 172.1~keV, and 375.0~keV, with branching fractions of 68.7\%, 25.7\%, and 17.3\% respectively~\cite{Hashizume2011}) that can escape the TPC and be detected in the Skin. The fraction of these decays that produce a coincident Skin signal provides a direct, data-driven measure of the veto efficiency for gamma rays in this energy range. Using WS2022 data, the combined, tagged, and untagged energy spectra are fit simultaneously via binned $\chi^2$ minimization (Fig.~\ref{fig:xe127_tagging}), with the $^{127}$Xe tagging fraction as a free parameter. The signal model describes $^{127}$Xe as a Gaussian peak, while the background model includes contributions from $^{125}$I, $^{124}$Xe double electron capture, and a flat continuum. The $^{124}$Xe normalization and peak widths are constrained by a prior from the $^{124}$Xe 2$\nu$2EC measurement~\cite{aalbers2025two}.

The $^{124}$Xe KK capture fraction measured by LZ of $(64.8 \pm 5.3)\%$~\cite{aalbers2025two} is in some tension with the theoretical prediction of ${\sim}74\%$ from interacting shell model calculations~\cite{nictescu2024theoretical}. The theoretical values are adopted in the fit, and this discrepancy introduces a source of systematic uncertainty in the $^{124}$Xe background modeling. In the standard fiducial volume extending up to 4~cm away from the TPC wall, the untagged $^{127}$Xe component is subdominant relative to $^{124}$Xe and the flat background, making the tagging fraction sensitive to $^{124}$Xe modeling assumptions. 

To improve robustness, the selection is restricted to events within 3 to 6~cm of the wall, where the near-wall region enhances the proportion of single-scatter $^{127}$Xe events relative to spatially uniform backgrounds. Figure~\ref{fig:xe127_tagging} shows the simultaneous fit in this near-wall region. The fitted tagging fraction is stable when varying the standoff selection around the nominal 3--6~cm range, indicating that the result is robust against the choice of boundaries. In contrast, fits in the full fiducial volume show a systematic dependence on the assumed $^{124}$Xe branching fractions, as the additional volume contributes predominantly background events with negligible $^{127}$Xe signal. The 3--6~cm selection is therefore adopted as the primary measurement.

The resulting tagging fraction is $(88 \pm 2)\%$, consistent with a simulated veto efficiency of $(90.4 \pm 0.6)\%$ for $^{127}$Xe decays using the GEANT4-based BACCARAT package~\cite{agostinelli2003geant4, akerib2021simulations,Rushton:2024nft}. The simulations also indicate minimal position dependence within the TPC. We adopt the use of the measured value as representative of the full fiducial volume for gammas in the few hundred keV range that scatter once in the TPC.

\subsection{Case Study: $^{127}$Xe M-shell EC}

\begin{figure}
	\centering
    \includegraphics[width=0.9\linewidth]{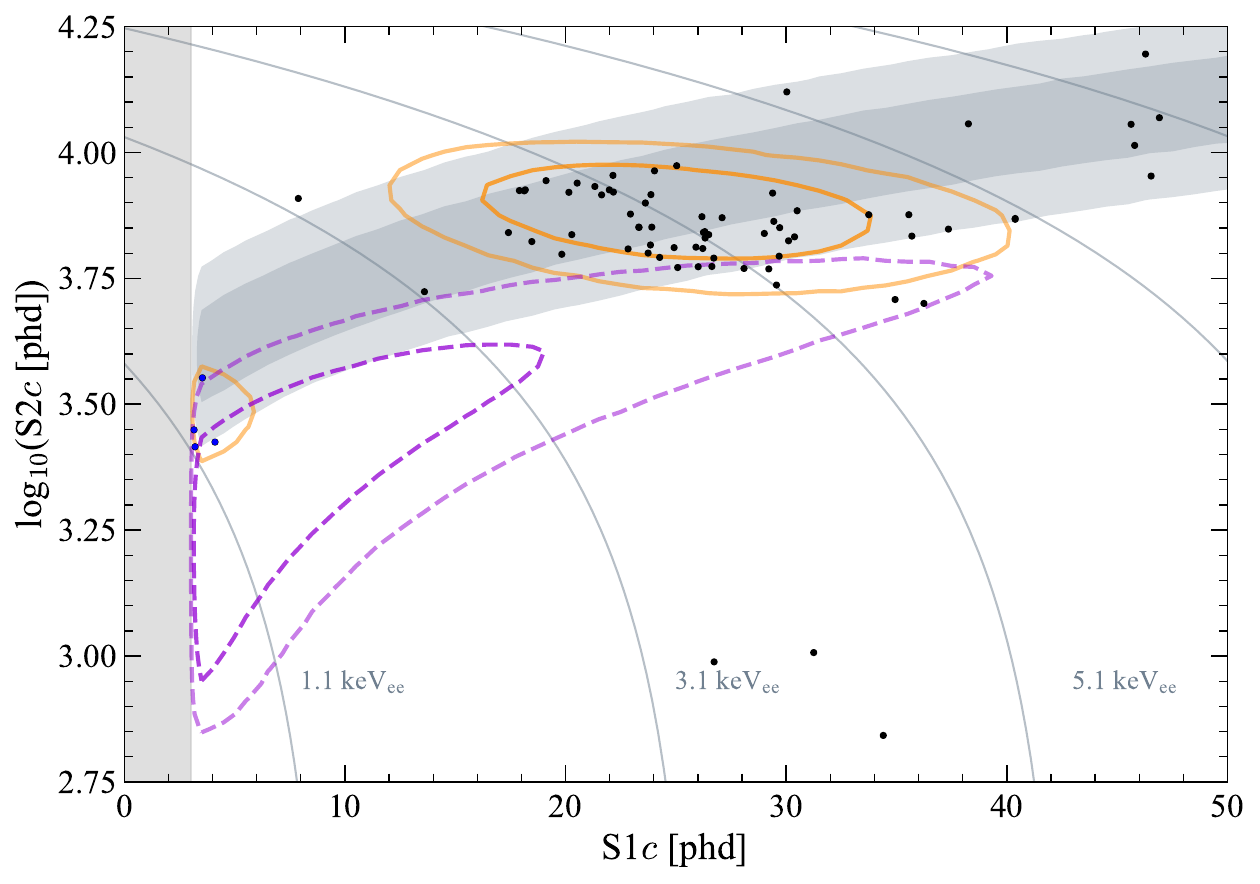}
	\caption{Distribution of WS2022 events with a coincidence Skin pulse in log(S2c)-S1c space, with a distance from 2 to 4 cm away from the wall. Four M-shell candidate events with 1.1~keV of energy are denoted by blue points, and cluster near S1c values of 3 or 4~phd. L-shell candidate events with 5.1~keV form the majority of the events at intermediate S1c values. The orange contours represent combined 1$\sigma$ and 2$\sigma$ $^{127}$Xe L/M-shell electron capture decays, where the disjoint contour at low energy corresponds to the M-shell. Other contours show the best-fit background model~\cite{aalbers2023first} (gray) and a 30~GeV/c$^{2}$ WIMP (purple dashed).}
    \label{fig:xe127_mshell}
\end{figure}

As a case study to illustrate how the Skin plays a role in understanding the detector, $^{127}$Xe M-shell EC events are used as a check on the signal efficiency of the TPC. In WS2022, LZ observed a deficit of events in the lower half of the ER band~\cite{aalbers2023first}, which led to stronger-than-expected WIMP exclusion limits. One hypothesis was that our energy threshold had systematically increased or fluctuated during the run, which would raise concerns about the validity of the calculated exclusion limits. To address this issue, the Skin was used to tag $^{127}$Xe decays that can be utilized as an internal calibration source. The M-shell electron capture decays of $^{127}$Xe deposit a total of 1.1 keV in X-rays and Auger electrons, the energy region where the deficit was observed.

Applying the same coincidence and timing criteria as in the Skin prompt veto to a modified fiducial volume of 2--4~cm away from the wall, four M-shell candidate events were identified. Figure~\ref{fig:xe127_mshell} shows the distribution of these events in log(S2c)-S1c space, along with background and signal model contours. The expected number of M-shell events is 3.2$\pm$0.6, calculated from the measured L-shell capture rate, branching ratios, recombination enhancement~\cite{aalbers2025measurements}, and the ER threshold~\cite{aalbers2023search}. With the statistical accuracy afforded by these small numbers, this measurement confirms that LZ maintained sensitivity throughout WS2022, supporting the interpretation that the observed deficit in the ER band was a statistical fluctuation rather than an instrumental effect.

\section{Summary}\label{sec:Summary}

This paper characterizes the performance of the Skin Detector based on data from different sources, focusing on energy response, spatial resolution, and event tagging capabilities.

The energy scale and vertical position reconstruction in the Skin Barrel are calibrated using monoenergetic gamma-ray sources deployed through the external calibration system. A position-dependent correction based on the top-bottom asymmetry of scintillation signals is developed to mitigate geometric non-uniformities. The resulting energy and spatial reconstruction provide a basis to assess prompt and delayed veto thresholds.

The alpha decays from $^{222}$Rn and $^{220}$Rn chains are used to measure spatial resolution, yielding $\sim$14 mm precision in both vertical and azimuthal directions. Time-correlated decay pairs are used to reveal stable convective flow structures within the detector, with mean vertical flow velocities on the order of 2--4~mm/s. These measurements offer valuable insight into the long-term fluid dynamics of large-scale LXe systems and can be integrated with more comprehensive flow studies in the TPC.

The Skin veto system demonstrates performance largely exceeding design goals, with prompt veto thresholds below 10~keV$_\text{ee}$ in the Barrel and below 100~keV$_\text{ee}$ in the Dome. The prompt veto efficiency for gamma rays is measured to be $(88 \pm 2)\%$ using Skin-tagged $^{127}$Xe K-shell decays. Reliable observation of low-energy events, including $^{127}$Xe M-shell electron captures, further validates the LZ detector's sensitivity in the WIMP search region.

\section*{Acknowledgements}


The research supporting this work took place in part at the Sanford Underground Research Facility (SURF) in Lead, South Dakota. Funding for this work is supported by the U.S. Department of Energy, Office of Science, Office of High Energy Physics under Contract Numbers DE-AC02-05CH11231, DE-AC02-76SF00515, DE-AC52-07NA27344, DE-SC0008475, DE-SC0010004, DE-SC0010010, DE-SC0010072, DE-SC0011702, DE-SC0012447, DE-SC0012704, DE-SC0014223, DE-SC0015535, DE-SC0015910, DE-SC0018982, DE-SC0019066, DE-SC0019193, DE-SC0020216, DE-SC0025446, DE-SC0025629, DE-SC0026014, DE-SC0026544. This research was also supported by U.S. National Science Foundation (NSF) ; the UKRI’s Science \& Technology Facilities Council under award numbers ST/M003639/1, ST/R003181/1, ST/W000466/1, ST/W000490/1, ST/W000555/1, ST/W00058X/1, ST/W000628/1, ST/W000636/1, ST/X00600X/1, ST/X006050/1, UKRI2837, UKRI2840, UKRI2841, UKRI2842, UKRI2843, UKRI2844, UKRI2847, UKRI2851, UKRI2852; the Portuguese Foundation for Science and Technology (FCT) under award number PTDC/FIS-PAR/2831/2020; the Institute for Basic Science, Korea (budget number IBS-R016-D1); the Swiss National Science Foundation (SNSF) under award number 10001549; the European Union’s research and innovation programme Horizon Europe under the Marie Skłodowska-Curie grant agreement No 101209429. This research was supported by the Australian Government through the Australian Research Council Centre of Excellence for Dark Matter Particle Physics under award number CE200100008. We acknowledge additional support from the UK Science \& Technology Facilities Council (STFC) for PhD studentships and the STFC Boulby Underground Laboratory in the U.K., the GridPP~\cite{Faulkner_2006,Britton_2009} and IRIS Collaborations, in particular at Imperial College London and additional support by the University College London (UCL) Cosmoparticle Initiative, and the University of Zurich. We acknowledge additional support from the Center for the Fundamental Physics of the Universe, Brown University. K.T. Lesko acknowledges the support of the Royal Society, the Wolfson Foundation, and Brasenose College. This research used resources of the National Energy Research Scientific Computing Center, a DOE Office of Science User Facility supported by the Office of Science of the U.S. Department of Energy under Contract No. DE-AC02-05CH11231. We gratefully acknowledge support from GitLab through its GitLab for Education Program. Fermilab is managed by Fermi Forward Discovery Group, LLC, acting under Contract No. 89243024CSC000002. The University of Edinburgh is a charitable body, registered in Scotland, with the registration number SC005336. The assistance of SURF and its personnel in providing physical access and general logistical and technical support is acknowledged. We acknowledge the South Dakota Governor's office, the South Dakota Community Foundation, the South Dakota State University Foundation, and the University of South Dakota Foundation for use of xenon. We also acknowledge the University of Alabama for providing xenon. For the purpose of open access, the authors have applied a Creative Commons Attribution (CC BY) license to any Author Accepted Manuscript version arising from this submission. The preparation of this manuscript was assisted by large language models, including a custom retrieval-augmented generation system ("LZ Oracle") for information retrieval and factual cross-checking, and general-purpose models (GPT-4o and Claude Sonnet 4.5) for language polishing and proofreading. The authors take full responsibility for the accuracy and correctness of the content. Finally, we respectfully acknowledge that we are on the traditional land of Indigenous American peoples and honor their rich cultural heritage and enduring contributions. Their deep connection to this land and their resilience and wisdom continue to inspire and enrich our community. We commit to learning from and supporting their effort as original stewards of this land and to preserve their cultures and rights for a more inclusive and sustainable future.

\bibliography{Skin-refs-v2.bib}

\end{document}